\documentclass[reprint,amsmath,superscriptaddress,amssymb,aps,twocolumn,prstab]{revtex4-1}
\usepackage{graphicx,subfigure,bm,amssymb,amsmath,dcolumn,hyperref}
\usepackage{color,multirow}
\usepackage{mathrsfs}
\usepackage{enumitem}

\begin{document}
\newcommand{\blue}[1]{\textcolor{blue}{#1}}
\newcommand{\red}[1]{\textcolor{red}{#1}}
\newcommand{\green}[1]{\textcolor{green}{#1}}
\newcommand{\orange}[1]{\textcolor{orange}{#1}}

\newcommand{\Lm}{L_{\rm min}}
\newcommand{\Vm}{V_{\rm min}}
\newcommand{\scrO}{\mathcal{O}}
\newcommand{\scrL}{\mathcal{L}}
\newcommand{\scrH}{\mathcal{H}}
\newcommand{\scrZ}{\mathcal{Z}}
\newcommand{\scrG}{\mathcal{G}}
\newcommand{\scrV}{\mathcal{V}}
\newcommand{\scrE}{\mathcal{E}}
\newcommand{\scrA}{\mathcal{A}}
\newcommand{\scrF}{\mathcal{F}}
\newcommand{\scrB}{\mathcal{B}}
\newcommand{\Clone}{F_1}
\newcommand{\Cltwo}{F_2}

\newcommand{\ZZ}{\mathbb{Z}}
\newcommand{\EE}{\mathbb{E}}

\newcommand{\yt}{y_t}
\newcommand{\yh}{y_h}

\title{Geometric properties of the complete-graph Ising model in the loop representation}
\author{Zhiyi Li}
\affiliation{Department of Modern Physics, University of Science and Technology of China, Hefei, Anhui 230026, China}	
\author{ZongZheng Zhou} 
\affiliation{ARC Centre of Excellence for Mathematical and Statistical Frontiers (ACEMS),
School of Mathematics, Monash University, Clayton, Victoria 3800, Australia}
\author{Sheng Fang}
\email{fs4008@mail.ustc.edu.cn}
\affiliation{Hefei National Research Center for Physical Sciences at the Microscale,
University of Science and Technology of China, Hefei 230026, China}
\affiliation{MinJiang Collaborative Center for Theoretical Physics,
College of Physics and Electronic Information Engineering, Minjiang University, Fuzhou 350108, China}
\author{Youjin Deng}
\email{yjdeng@ustc.edu.cn}
\affiliation{Department of Modern Physics, University of Science and Technology of China, Hefei, Anhui 230026, China}
\affiliation{Hefei National Research Center for Physical Sciences at the Microscale,
University of Science and Technology of China, Hefei 230026, China}
\affiliation{MinJiang Collaborative Center for Theoretical Physics,
College of Physics and Electronic Information Engineering, Minjiang University, Fuzhou 350108, China}	

\begin{abstract}
The exact solution of the Ising model on the complete graph (CG) provides an important, though mean-field, insight for the theory of continuous phase transitions. 
Besides the original spin, the Ising model can be formulated in the Fortuin-Kasteleyn random-cluster and the loop representation,
in which many geometric quantities have no correspondence in the spin representations. 
Using a lifted-worm irreversible algorithm, we study the CG-Ising model in the loop representation, and, 
based on theoretical and numerical analyses, obtain a number of exact results including volume fractal dimensions and scaling forms. 
Moreover, by combining with the Loop-Cluster algorithm, we demonstrate how the loop representation can provide an intuitive understanding to 
the recently observed rich geometric phenomena in the random-cluster representation, 
including the emergence of two configuration sectors, two length scales and two scaling windows. 
\end{abstract}
\date{\today}
\maketitle

\section{Introduction}
\label{SecI}

% The Ising model~\cite{duminil2022100} is one of the most important model of statistical physics which helps us better study the phase transition and critical phenomenon. It describes the system of ferromagnetic with spin interaction in the nearest neighbour. On the other hand, Landau theory \cite{landau1937theory} is a general mean-field theory of continuous phase transition. It provides an exceptionally broad and powerful framework for the phase transition theory with the associated concept of the order parameter as an essential character and the process of spontaneous symmetry breaking \cite{kardar2007statistical}. Ising model on the complete graph (CG) is one of the exactly solved models in statistical mechanics, which is always referred to the mean-field model in a sense ``infinite-dimensional" \cite{baxter2016exactly}. On the CG, each spin interacts equally with every other spin. With a variable transformation from local spin to global magnetization and a quartic perturbation approximation \cite{luijten1997interaction,deng2004constrained}, the free energy of the CG Ising model can be converted to the form of interaction analogous to the free energy in Landau theory. Therefore, the study of the Ising model on CG can provide a significant understanding for the Landau mean-field theory of continuous phase transition. \par

The Ising model~\cite{duminil2022100} is one of the most prototypical models in statistical physics
and plays an important role in the study of phase transitions and critical phenomena.
It has wide applications in many fields, including material science,  neuroscience and biology, etc. 
Given a graph (or lattice) $\scrG=(\scrV,\scrE)$ with the vertex set $\scrV$ and edge set $\scrE$, the Hamiltonian of the zero-field ferromagnetic Ising model reads
%It describes the ferromagnetic system with spin interaction in the nearest neighbor, and the Hamiltonian is 
%\begin{equation}
%\label{eq:model_spin}
%\scrZ_{\rm{spin}} = \sum_{\{s\}} e^{-\beta \scrH(\{s\}) } = \sum_{\{s\}} \prod_{\langle ij \rangle}  e^{\beta J s_i s_j }, 
%\end{equation}
\begin{equation}
\label{eq:model_spin}
\scrH(s) = - J \sum_{ij \in \scrE} s_i s_j\;, 
\end{equation}
where $J>0$ is the interaction strength. The probability of a spin configuration $s \in \{-1,1\}^{\scrV}$ is given by the Gibbs measure $\pi(s) \propto e^{-\beta H(s)}$, where $\beta$ is the inverse temperature. Let $K :=\beta J$ be the reduced coupling strength, and one can set $J=1$ for convenience. On lattices $\ZZ^d$, it has been rigorously established that the Ising model goes through a continuous phase transition for $d\ge2$~\cite{Onsager1944Crystal,AizenmanFernandez1986,Aizeman2015Random}.
%The Ising model undergoes  a continuous phase transition for spatial dimensions $d\ge2$, and in most cases, it cannot be solved exactly.  
%As a qualitative description, the mean-field theory treats the interaction as an external field and provides an exceptionally broad and powerful framework for the phase transition theory~\cite{kardar2007statistical}.
%with the associated concept of the order parameter as an essential character and the process of spontaneous symmetry breaking~\cite{kardar2007statistical}. 
%A candidate for mean-field approach, which is particularly insightful for finite-size scaling (FSS) behaviors, is to study the Ising model 
%An example of the mean-field model is the Ising spins on the complete graph (CG), where  each vertex is connected to all others, i.e., the total edge number $|\scrE|= \frac{V(V-1)}{2}$ and the volume $V = |\scrV|$. 
%Thus, every spin is identical and the interaction from all other spins can be regarded as the external field. 
%For the Ising model on the CG, the reduced coupling strength is set to be $K/V$ to satisfy the extensive property of statistical systems.
 \par 

% Originally, the zero-field Ising model is described by the spins of the lattice system, with the probability measure of each spin configuration on $V$-vertices CG written as
% \begin{equation}
%     \pi(\{s\})\propto\exp{[\frac{K}{V}\sum_i\sum_{i\neq j}s_is_j]},
%     \label{spin_par}
% \end{equation}
% where $s_i$ is the spin of the $i$-th vertex, $K$ is the coupling constant of spin. 
%The Ising model has been extensively studied, mostly in its spin representation. 
In addition to its spin representation in Eq.~(\ref{eq:model_spin}), the Ising model can be formulated in two other geometric representations,
the loop representation and the Fortuin-Kasteleyn (FK) bond representation.
Here, we provide a brief overview of these two representations for clarity. 
% Here, we demonstrate these two geometric representation of the Ising model on the CG.
In 1941, Van der Waerden proposed a high-temperature expansion trick~\cite{van1941lange} for the Ising model,
where the statistical weight for each interaction term is rewritten as 
$\exp (Ks_is_j)=\cosh{K} (1+s_is_j \tanh{K})$. 
% Further, to geometrically represent the product $\prod$ in Eq.~(\ref{eq:model_spin}), 
Further, an auxilliary variable $f_{ij} = 0, 1$ is introduced
such that the second term, $s_i s_j \tanh K$, is geometrically represented by an occupied bond $f_{ij}=1$,
and the first term corresponds to an empty bond $f_{ij}=0$. 
Then, the spin degrees of freedom can be integrated out by calculating 
the partition function $\scrZ_{\rm{spin}} = \sum_{s} e^{-\beta \scrH(s) }$, leading to the summation of geometric configurations of bond variables $f$.
Due to the $\mathbb{Z}_2$ symmetry of the Ising spins, 
non-zero contributions to the partition function come only from those configurations $\scrF $, 
in which any vertex is incident to an even number of occupied bonds. 
Such a configuration is composed of loops (also called currents or flows).
In graph theory, such a loop configuration is referred to an Eulerian graph or an even graph.
Let ${\rm even}(\scrG)$ be the set of loop configurations on $\scrG$. Then loop Ising model is defined by giving 
%This leads to the loop representation~\cite{HD2016RC},
any $\scrF$ the probability measure
%From it, one defines the loop representation of the Ising mode, which is named after that the configuration is 
%made up of a soup of loops~\cite{HD2016RC}.
%Any vertex on a loop configuration is incident to an even number of bond, so in graph theory, 
%we also refer the loop configuration as an Eulerian graph or an even graph. 
%Therefore, the probability measure of any configuration $\scrF$ on it is written as
 \begin{equation}
\pi(\scrF)\propto w^{|\scrF|}\delta_{\scrF\in {\rm even}(\scrG)}\; ,
\label{eq:model_loop}
\end{equation}
%\delta_{\scrF\in {\rm \emph{even}}(\scrG)}
where $|\scrF|$ represents the total number of occupied bonds, 
the bond weight is $w=\tanh{K}$ and $\delta_{\scrF\in {\rm even}(\scrG)}$ is an indicator function that ensures that any graph $\scrF$ descried by the flow variables is an even graph.
Apart from the Eulerian requirement, the probability measure~(\ref{eq:model_loop}) would describe 
the standard bond percolation, and, thus, the loop representation of the Ising model 
can be regarded as the Eulerian bond percolation model. 
Other names for this representation include the random-current model\cite{HD2016RC}, random even graph\cite{GrimmiteRandomEvenGraph} or the flow representation of the Ising model.

 %In contrast to the phase transition in the random-cluster representation on the CG, the critical properties (e.g. the fractal dimension and other scaling behaviors) in the loop representation has not been widely studied yet. \par 
The $Q$-state Potts model~\cite{Wu1982}, in which the value of spins can take $\sigma \in \{0, 1, \cdots, Q-1\}$, is a generalization of the Ising model 
and has the latter as a special case of $Q=2$.
In 1969, Fortuin and Kasteleyn established an exact mapping between the Potts model and 
a geometric model, called the random-cluster (RC) model~\cite{fortuin1972random,grimmett2006random}. % The $Q=2$ RC model is called the Fortuin-Kasteleyn (FK) random-cluster representation for the Ising model.
Similar to loop configuration, for each edge $ij$, a binary variable $b_{ij} \in \{0,1\}$ is defined to represent whether the edge is occupied by a bond ($b_{ij} = 1$) or empty ($b_{ij} = 0$), but no Eulerian constraint is required in the FK configurations. 
%Each connected component (also called \emph{cluster}) has a weight $Q$, 
The $Q$-state RC model is defined by choosing a spanning subgraph $\scrA \subseteq \scrG$ with the probability
\begin{equation}
\pi(\scrA)\propto Q^{k(\scrA)}p^{|\scrA|}(1-p)^{|\scrE|-|\scrA|},
\end{equation} 
where $p$ is the bond occupation probability and $k(\scrA)$ is the number of connected components (or \emph{clusters}) on $\scrA$. 
The case $Q=2$ with $p=1- e^{-2K}$ corresponds to the Ising model, or known as the FK Ising model, where $K$ is the reduced coupling strength mentioned before.\par
\begin{figure}[t]
    \centering
    \includegraphics[width=0.5\textwidth]{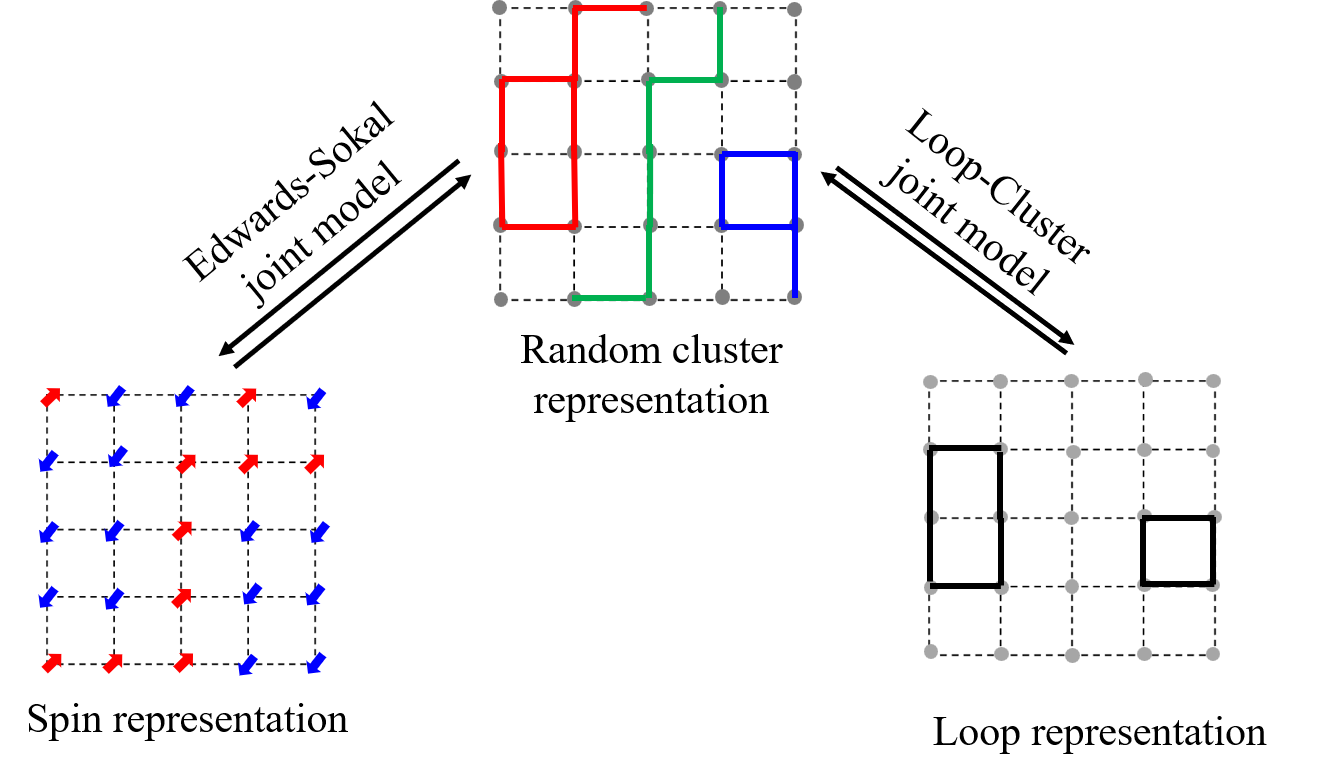}
    \caption{Three representations of the Ising model. The random-cluster representation is depicted with lines of different colors indicating different clusters. The Edwards-Sokal joint model couples the spin and random-cluster representations, while the Loop-Cluster joint model couples the loop and random-cluster representations.}
    \label{fig:Graph_rep}
\end{figure}

These three representations are illustrated in Fig.~\ref{fig:Graph_rep}.
%For $q=2$, it reduces to the zero-field ferromagnetic Ising model with $p=1-e^{-2K/V}$. 
In comparison to the spin representation, geometric representations enable the definition of a broader range of geometric observables, many of which have no corresponding analogs in the spin representation, leading to a wealth of phenomena. For example, while the upper critical dimension of the spin Ising model has been known to be $d_c = 4$ since the 1970s, recent studies argued that the FK-Ising model simultaneously has two upper critical dimensions, namely $d_c=4$ and $d_p=6$ \cite{Fang_2022,Fang_2022_2}, where the dimension $d_p = 6$ cannot be observed in the spin representation. Moreover, the geometric representations also serve as a versatile platform for conformal field theory~\cite{francesco2012conformal} and stochastic Loewner evolution~\cite{kager2004guide,cardy2005sle}, leading to many exact results in two dimensions. In mathematical physics, the geometric representations play a crucial role in the rigorous study of phase transitions for the Ising model in dimensions $d \geq 3$~\cite{AizenmanFernandez1986,Aizeman2015Random} and 
the triviality of criticality for $d = 4$~\cite{Aizeman2021Marginal}. 

Advanced Monte Carlo methods have also benefited from the geometric representations. 
A notable example is the highly efficient Swendsen-Wang cluster algorithm~\cite{PhysRevLett.58.86}, 
passing back and forth between the FK and the spin configurations.
The Edwards-Sokal joint model~\cite{PhysRevD.38.2009}, in which the spin and FK-bond variables are coupled together, 
establishes a connection between the FK and the spin representations and offers a concise understanding 
to the Swendsen-Wang algorithm. 

Another example is the Loop-Cluster (LC) algorithm, which passes back and forth between the FK and loop representations 
via the LC joint model~\cite{Zhang_2020}.  
% which passes back and forward between the FK and the loop representation.  Recently, another joint model with the random-cluster representation 
% and the loop representation, named Loop-Cluster(LC) joint model, 
% was generated inspired by the Edward-Sokal model~\cite{Zhang_2020,GrimmiteRandomEvenGraph}. 
 A configuration of the LC joint model can be interpreted as a superposition of a FK and a loop configuration, 
 where each edge is associated with both a FK bond and a flow variable. 
 For the Ising model ($Q=2$), the probability measure of an LC joint configuration is defined as follows,
 \begin{equation}
\pi(\scrA,\scrF)\propto \left(\frac{p}{2}\right)^{|\scrF|}\left(\frac{p}{2}\right)^{|\scrA|-|\scrF|}(1-p)^{|\scrE|-|\scrA|}\delta_{\scrF\in {\rm even}(\scrG)}.
\label{jointLC}
 \end{equation}
More specifically, there are four edge states in the LC joint model:
\begin{align}
    \scrL_1:b_{ij}=0, f_{ij}=0; \notag \\
    \scrL_2:b_{ij}=1, f_{ij}=0; \notag \\
    \scrL_3:b_{ij}=1, f_{ij}=1; \notag \\
    \scrL_4:b_{ij}=0, f_{ij}=1 .
    \label{edgestate}
\end{align}
The edge state $\scrL_4$, i.e, with an empty FK bond and an occupied loop bond, is forbidden. From Eq.~\eqref{jointLC} the probabilities of other edge states read:
\begin{equation}
    P(\scrL_1) = 1-p;\ \ P(\scrL_2)=\frac{p}{2};\ \ P(\scrL_3)=\frac{p}{2}.
\end{equation}
% The LC joint model is applied to the design of a new Monte Carlo (MC) method called Loop-Cluster algorithm, 
% realizing the transformation between the two representation. 
From a given loop configuration $\{f \}$, the LC algorithm generates a stochastic FK bond configuration $\{b \}$ 
by a local bond-placing process. To be specific, for each nonzero flow $f_{ij} = 1 $, one sets $b_{ij} = 1 $; 
for each empty flow $f_{ij} = 0$, one independently sets $ b_{ij} = 1 $ with probability
\begin{equation}
    w^\prime = \frac{\frac{p}{2}}{\frac{p}{2}+(1-p)} = \tanh{K},
\end{equation}
which equals to $w$, or $b_{ij} = 0$, otherwise. 
In other words, the process from the loop to the FK representation is basically to add occupied bonds to 
those edges of empty flow via a process of standard bond percolation with probability $w$. \par
In statistical mechanics, it is of particular interest to study models on the complete graph (CG) because it is usually more tractable and provides important insights to understand critical behaviors on high-dimensional tori.
% In the field of statistical mechanics, the complete graph (CG) serves as a valuable tool for investigating models due to its tractability and ability to shed light on critical behaviors in high-dimensional finite systems. 
On the CG, each vertex is connected to all others, and thus in order to obtain an extensive system, the coupling strength $K$ must be rescaled by its volume $V$.
%The CG-Ising model in the spin representation has been solved exactly, where the critical coupling strength is at $K_c/V =1/V$.
%FSS behaviors of a variety of quantities can be found in Ref.~\cite{luijten1997interaction}. 
%Thus, the critical bond fugacity are  $w_c=\tanh{1/V} \approx 1/V$ and $p_c = 1-e^{-2/V} \approx 2/V$ for the loop representation and FK representation, respectively.
For the FK-Ising model on the CG, it was proven that~\cite{luczak2006phase}, within a scaling window of width $O(V^{-1/2})$ 
around the critical point $p_c = 1-e^{-2/V}$, the sizes of the largest and second-largest clusters scale asymptotically 
as $C_1 \sim V^{3/4}$ and $C_2 \sim \sqrt{V} \ln V$, respectively.
Namely, unlike the common self-similarity observed in many other critical systems, here $C_1$ dominates over $C_2$, which indicates the system has two length (size) scales~\footnote{Spatial length is not defined on the complete graph. Here the two length (size) scales mean that, compared with the self-similarity property commonly observed in many critical systems, the largest cluster in the FK Ising model is much larger than other clusters.}.
% both at criticality and in its critical window with width $O(\sqrt{V})$, 
% which indicates the appearance of two-length scales.
Furthermore, the authors also proved that $C_1, C_2 \sim V^{2/3}$ in a wider scaling window of width $\delta p \equiv p_c-p = O(V^{-1/3})$ 
in the sub-critical side ($\delta p < 0$), sharing the same scaling behavior as the CG-\emph{percolation} model. Thus, the FK-Ising model on the CG has two scaling windows. Additionally, the authors in Ref.~\cite{PhysRevE.103.012102} numerically studied the FK-Ising model on the CG and observed more interesting critical phenomena. 
At criticality, the cluster-number density of the FK-Ising clusters, excluding the largest one, 
obeys the same scaling form as that for the bond percolation on the CG.
Moreover, a percolation sector was observed in the whole configuration space, 
which asymptotically vanishes with the rate of $V^{-1/12}$.
Conditioned on being in the percolation sector, all clusters, including the largest one, have the same scaling behavior 
as those for the critical percolation on the CG.\par 

% Inspired by the rich phenomena in the FK representation, i.e., the emergence of the two configuration sectors, two length scales and two scaling windows,  
% one would wonder whether such rich scaling behaviors can be observed in the loop representation, 
% since the two geometric representations can be linked via the LC joint model. 
% From the LC joint model, we are expecting to gain some insights on the observation of rich critical phenomena in the random-cluster representation on the CG. Theoretically, it is proved that the random cluster representation exhibits two critical windows\cite{luczak2006phase}, as size of the largest cluster scales as $C_1\sim V^{3/4}$ within the Ising critical window given by $1-p/p_c = {O}(\frac{1}{\sqrt{V}})$ while it scales as $C_1\sim V^{2/3}$ within the percolation critical window given by $1-p/p_c = {O}(V^{-1/3})$, with the critical point $p_c = \frac{2}{V}$. Recently, it's numerically discovered that the bond configuration space of the FK representation can be partitioned into two sectors: the percolation sector $S^P$ and the Ising sector $S^I$ according to the size of the largest cluster while $S^P$ is vanishing as $V^{-1/12}$. Besides, the scaling behavior of the first- and second- largest clusters are different as $C_1^{\rm FK}\sim V^{3/4}$ but $C_2^{\rm FK}\sim \sqrt{V}\log{V}$.\par
In this paper, we study the CG-Ising model in the loop representation by a lifted worm update algorithm \cite{PhysRevE.97.042126}. 
The motivation is two-fold. 
Firstly, we aim to examine the critical scaling behaviors of the geometric clusters in the loop representation.
Secondly, given that the process in the LC algorithm from the loop to the FK bond configurations is much like 
the conventional percolation process, we hope to gain a vivid understanding of 
the observed rich geometric properties for the CG-Ising model in the FK representation. 

At criticality, we first study the number of occupied loop bonds (i.e., nonzero flows) $\scrB$. Based on the exact solution of the spin Ising model on the CG, we derive the mean scales as $B := \langle\scrB\rangle \asymp \sqrt{3}\frac{\Gamma(3/4)}{\Gamma(1/4)} \sqrt{V}$\footnote{In this paper, $A_V \asymp B_V$ means that $\lim_{V\rightarrow \infty}\frac{A_V}{B_V} = 1$ while $A_V \sim B_V$ means their ratio converges to some positive constant.}, and the probability density function of $X_B\equiv \scrB/\sqrt{V}$ is,
\begin{equation}
    f_{X_B}(x) \asymp \frac{3^{-1/4}}{2}\Gamma^{-1}(5/4) x^{-1/2}\exp\left(-\frac{1}{3}x^2\right)\;,
    \label{eq:dist_flowbond}
\end{equation}
which is also verified by our numerics. It means that, in contrast to the FK-Ising model, the number of bonds in the loop Ising model is not extensive and has a power-law distribution till $O(\sqrt{V})$. % Besides, it is interesting to observe that the probability for a fully vacant configuration, i.e., all the flow variables are zero, algebraically decays as $V^{-1/4}$.
Meanwhile, through the results of our simulations, we conjecture that the total number of flow clusters increases logarithmically as $\frac{1}{4}\ln V$, and each flow cluster is basically unicyclic. 
In other words, a typical loop configuration consists of an extremely dilute soup of cycles. 
The cluster-number density $n(s,V)$ of flow clusters, including the largest one, obeys the scaling form as 
\begin{equation} 
n(s,V) \asymp \frac{1}{2V} \; s^{-1} \tilde{n} (s/\sqrt{V})  \hspace{5mm} \mbox{with  } \; \tilde{n}(x \to 0)=1. 
\label{eq:flow_ns}
\end{equation}
The sizes of the largest and second-largest flow clusters both scale 
as $ F_1, F_2 \sim \sqrt{V}$, and, accordingly, we conjecture that the volume fractal dimension~\footnote{The volume fractal dimension is to characterize how cluster sizes scale with respect to the system volume.} $d_{\rm f1} = d_{\rm f2} =1/2$ 
holds exactly true. 
Unlike in the FK representation, the size distribution of the largest flow cluster 
displays a power-law behavior until the cut-off size $O(\sqrt{V})$, 
and its scaling form in the rescaled variable $ X_1\equiv \scrF_1/\sqrt{V}$ reads 
\begin{equation} 
f_{X_1}(x) \asymp \frac{1}{2} \, x^{-1/2}\tilde{f} (x)  , 
\label{eq:flow_dist}
\end{equation}
where function $\tilde{f} (x \to 0)=1$ and $\tilde{f}$ drops quickly for $x \gg 1$.
% Actually, the total number of nonzero flows (i.e., occupied loop bonds) is not extensive and scales as $\sqrt{V}$, 
% and the corresponding flow density also exhibits a power-law distribution, following Eq.~(\ref{eq:flow_dist}). 
% It is interesting to observe that the probabilty for a vacant configuration, 
% i.e., all the flow variables are zero, algebraically decays as $V^{-1/4}$.
% where the value of the exponent might hold exactly true. 
% Near the criticality with $\delta p = p_c-p$, $B$ are observed to follow the conventional FSS ansatz as 
% ${\cal Q} \sim V^{y_{\cal Q}} \tilde{\cal Q} (\delta p \sqrt{V})$, 
% where $y_{\cal Q}$ is the corresponding exponent and only a single scaling window of width $\scrO(\frac{1}{\sqrt{V}})$ appears. 
Near the criticality with $\delta p$, $B$ can be demonstrated to follow the conventional finite size scaling (FSS) ansatz as 
$B = \sqrt{V} \tilde{B}(\delta p\sqrt{V})$ with $\tilde{B}(\cdot)$ the scaling function, and only a single scaling window of width $O(1/\sqrt{V})$ appears.\par
Therefore, in the loop representation, no apparent symptoms are observed for the appearance of the two length scales, 
of two configuration sectors, and of two scaling windows, which occur in the FK representation of the CG-Ising model. However, the loop representation provides a starting point for us to understand these rich phenomena with the LC algorithm. The density of bonds $B/V$ scales as $1/\sqrt{V}$, which suggests that the loop configurations are very dilute and become vacant as $V\to \infty$. Further, in the LC process, the probability of adding bonds to the loop configuration on the CG is $w_c = \tanh(1/V) \approx 1/V$, which is equal to that for the bond percolation process at criticality. Overall, the LC process can be roughly viewed as the critical bond percolation process on the CG.
% Note that the forms of scaling functions (\ref{eq:flow_dist}) and (\ref{eq:flow_ns}) 
% and the exact values of several exponents are merely conjectured 

% by FSS (FSS) ansatz,
% We conjecture the fractal dimension of the loop representation as $d_{\rm F}^l = \frac{1}{2}$ from the result of the size of the largest loop cluster. 
% Besides, we also observe that the size of the largest and second-largest loop clusters both scale as $\sqrt{V}$, 
% in contrast to the different scaling length of the random-cluster representation. We also discover that in the loop representation, the number of bonds and clusters are not extensive, which means each vertex is isolated in the thermodynamic limit.
% Another interesting fact we observe is that
% each cluster in the loop representation is uni-cyclic as the number of bonds and the number of vertices converge to the same value.  
% From our result of the cluster-size distribution $n(s,V)$ and the probability distribution of the largest loop cluster $f_{\mathcal{L}_1}(s,V)$, 
% we find that the configuration space of the loop representation is neat without more scaling sector.\par
More specifically, using the LC algorithm, we numerically find that the fraction of the loop bonds in the largest FK cluster tends to $1$, which means all the loop bonds belong to the largest FK cluster in the thermodynamic limit. In other words, all the other FK clusters are indeed generated by the critical percolation process in the LC algorithm. As a consequence, the emergence of two length scales in the critical FK configurations can be understood straightforwardly. % The bond-adding process in the LC algorithm is identical to the standard bond percolation process with the critical occupation probability $p_c=1/V$.
Almost all loops are merged together by the newly added FK bonds, leading to a giant cluster with the volume fractal dimension $D_{\rm f1}=3/4$. The remaining clusters are effectively generated by adding bonds on the vacant space, and thus, they behave like those percolation clusters on the CGs. \par
It can be calculated from Eq.~\eqref{eq:dist_flowbond} that the probability of the vacant configuration in the loop representation scales as $V^{-1/4}$, it follows that there must exist a percolation sector decaying slowly as or more slowly than the order $V^{-1/4}$ in the FK representation. Since the volume fractal dimension of the cycle (bridge-free) in the CG percolation model is $1/3$~\cite{PhysRevE.97.022107} at the critical point, we conjecture that if the size of the largest loop cluster $\scrF_1$ is no bigger than ${\cal O}(V^{1/3})$, the corresponding FK configurations belong to the percolation sector $S^{\rm P}$. We derive that $P(S^{\rm P})$  decays as $V^{-1/12}$, providing an explanation to the previous numerical observation~\cite{PhysRevE.103.012102} in the FK representation.
We then measure the scaling of the largest FK cluster $C_1^{\rm P'}$ conditioned on the original loop configuration with $\mathcal{F}_1\leq O(V^{1/3})$. Our data show that $C_1^{\rm P'}$ scales as $V^{2/3}$, which is the same as the CG-percolation. Moreover, from the scaling behavior of $B$ near the critical point, we find out that if $\delta p = O(V^{-1/3})$ then $B$ scales the same as the number of bridge-free bonds in the percolation model. Thus, it explains the two-scaling-window behavior of the FK representation. \par
Recall that the rich phenomena observed in Ref.~\cite{PhysRevE.103.012102} shows that there are strong percolation effects in the FK Ising model on the CG. These percolation effects now can be well understood from the perspective of the LC joint model.  It further reveals that configurations of the FK-Ising model excluding the largest cluster are effectively equivalent to the ones of percolation at the critical point on the CG. Meanwhile, the emergence of the percolation scaling windows in the FK representation suggests that when the temperature become higher, even the scaling behavior of the largest cluster is also described by percolation.
% \blue{
 % Since the bond density tends to 0 in the thermodynmaic limit, the whole graph is almostlty vacant. Morevoer, the probability of placing bond for the LC algorithm is $1/V$, consistent with the critcal thershold of the CG percolation. Thus, the LC algorithm performs the percolation process.    
% }

\par 
The remainder of this paper is organized as follows. Section \ref{secII} summarizes the simulation details and the sampled quantities.
Section \ref{secIII} presents our theoretical analysis. Section \ref{secIV} contains our main numerical results.  A discussion
is given in Sec.~\ref{secV}.
\section{Simulation \& observable}
\label{secII}
\subsection{Algorithm}
The worm algorithm~\cite{PhysRevLett.87.160601} is used to simulate the Ising model in the loop representation. The main idea of the worm algorithm is to enlarge the configuration space from close loop space to the space of the graph allowing two open ends by introducing two defects, i.e., vertices with odd degree. Configurations are updated as defects do random walks. If a defect proposes to move through a flow/loop bond, then with probability 1 the proposal is accepted and the bond is erased. If a defect is crossing an empty edge, then with probability $w$ the move is accepted and the empty edge is occupied by a bond. When two defects meet, a new loop configuration is obtained.\par

In Ref.~\cite{PhysRevE.97.042126}, the authors presented an irreversible version of the worm algorithm by using the lifting technique, which leads to a critical speeding-up for observables in the simulation on the CG, with a negative dynamic exponent $z=-1/2$. In other words, between two subsequent effectively independent samplings in the Markov chain, the number of elementary updating steps is of order ${\cal O}(\sqrt{V})$ in the lifted worm algorithm. This is vanishingly small in comparison with a sweep of updates, ${\cal O}(V)$, which is a standard unit in studying the efficiency of Monte Carlo methods. The existence of this critical speeding-up, which makes the lifted worm algorithm thus far most efficient for the CG-Ising, is understandable since the number of loop bonds is also ${\cal O}(\sqrt{V})$. 
Therefore, we use the irreversible worm algorithm to update loop configuration here. Specifically, a lifted parameter $\lambda\in\{+,-\}$ is introduced to double the configuration space of worm update, as $\lambda=+
(-)$ stands for the choice to add (delete) a bond in each step of random walk of the defect. It indicates that every time the defect moves to the next vertex, $\lambda$ determine whether the movement leads to an increase or decrease of bond on the graph, as well as the choice of the next vertex. Then we accept the update with a certain probability depending on $\lambda$ and the number of occupied bonds incident to the two defects, which is presented in Ref.~\cite{PhysRevE.97.042126} in details. Whenever the update is rejected, the lifted parameter $\lambda$ changes. \par
In addition, we implement a transformation from the loop representation to the RC representation via the LC algorithm~\cite{Zhang_2020} after we generate a loop configuration. The main idea of the transformation is performing a conditional probability distribution of the joint model \eqref{jointLC}. Recall that edge state $\scrL_4$ in Eq.~\eqref{edgestate} is forbidden, so a loop bond must also be an FK bond in the joint model.  Therefore, the basic step is that: for each edge, if it has not been occupied by a bond in the loop representation, we place a bond on it with a probability $w=\tanh{(K/V)} \approx K/V$; If it has been occupied, keep it occupied. We carry out this adding bond process by an efficient cumulative method~\cite{PhysRevE.66.066110}. 

\subsection{Sample quantities}
We sample the following observables in our simulations:
\begin{enumerate}[label= (\alph*)]
    \item The sizes of the largest and the second-largest loop clusters denoted as $\mathcal{F}_1$,  $\mathcal{F}_2$;
    \item The total number of vertices in the loop clusters $\mathcal{N}_{\rm v} = \sum_{i: \scrF_i > 1} \scrF_i$;
    \item The number of bonds $\scrB$ in loop clusters;
    \item The number of loop clusters $\mathcal{N}(s)$ with size $s$, defined as the number of loop clusters with size in $[s,s+\Delta s]$ with an appropriately chosen interval size $\Delta s$;
    \item The total number of loop clusters $\mathcal{N}_{\rm k}$;
    \item The indicators $\mathcal{P}^{\rm v}$,$\mathcal{P}^{(\alpha)}$. We set $\mathcal{P}^{\rm v}=1$ to record the event that the configuration is empty with bonds, $\mathcal{P}^{(\alpha)} =1$ to record if $\mathcal{F}_1\leq \alpha V^{\frac{1}{3}}$ with $\alpha$ is a tunable constant. Here we set $\alpha = 1,2$.
\end{enumerate}
% (a) the sizes of the largest and the second-largest loop clusters denoted as $\mathcal{F}_1$,  $\mathcal{F}_2$;\par
% (b) the number of bonds $\mathcal{N}_B$;\par
% (c) the number of loop clusters for each size $\mathcal{N}(s)$, defined as the number of loop clusters with size in $[s,s+\Delta s]$ with an appropriately chosen interval size $\Delta s$;\par
% (d) the total number of loop clusters $\mathcal{N}_{\rm k}$ ;\par
% (e) the total number of vertex in the loop clusters $\mathcal{N}_{\rm v}$\par
% (f) the indicators $\mathcal{P}^{n.v.}$,$\mathcal{P}^{(1)}$,$\mathcal{P}^{(2)}$ , as we set $\mathcal{P}^{n.v.}=1$ to record the event that the configuration is not empty with bonds,$\mathcal{P}^{(1)} =1$ to record if $\mathcal{F}_1\leq V^{\frac{1}{3}}$ while $\mathcal{P}^{(2)} =1$ to record if $\mathcal{F}_1\leq 2V^{\frac{1}{3}}$. \par
From these observables, we take the ensemble average:
\begin{enumerate}[label=(\alph*)]
    \item The probability of vacant configuration $P^{\rm v}=\langle\mathcal{P}^{\rm v}\rangle$;
    \item The average sizes of the first- and second-largest loop cluster $F_1=\langle\mathcal{F}_1\rangle,F_2=\langle\mathcal{F}_2\rangle$ and their distribution;% $f_{\mathcal {F}_1}(s)$;
    \item The average number of bonds $B=\langle\mathcal{B}\rangle$ and its distribution; %$f({\mathcal{B}})$;
    \item The average number of clusters $N_{\rm k}=\langle\mathcal{N}_{\rm k}\rangle$ and the average number of vertices $N_{\rm v}=\langle\mathcal{N}_{\rm v}\rangle$;
    \item The cluster-number density $n(s,V)=\frac{1}{V\Delta s}\langle\mathcal{N}(s)\rangle$, which is also called the cluster-size distribution;
    \item The probability of the bond configuration in the region where the largest loop-cluster $\mathcal{F}_1 \leq \alpha V^{1/3}$: $P(\mathcal{F}_1\leq \alpha V^{\frac{1}{3}})=\langle\mathcal{P}^{(\alpha)}\rangle$.
\end{enumerate}
% (a) the probability of vacant configuration $P^{v}=1-\langle\mathcal{P}^{n.v.}\rangle$;\par
% (b) the average sizes of the largest, the second-largest loop cluster $F_1=\langle\mathcal{F}_1\rangle$ and their distribution $f_{\mathcal {L}_1}(s),f_{\mathcal {L}_2}(s)$;\par
% (c) the average number of bonds $B=\frac{\langle\mathcal{N}_B\rangle}{V}$ and its distribution $f_{\mathcal{N}_B}(b)$;\par
% (d) the average number of clusters $N_{\rm k}=\langle\mathcal{N}_{\rm k}\rangle$ and the average number of vertices $N_{\rm v}=\langle\mathcal{N}_{\rm v}\rangle$;\par
% (e) the cluster size distribution $n(s,V)=\frac{1}{V\Delta s}\langle\mathcal{N}(s)\rangle$;\par
% (f) the probability of the bond configuration in the region where the largest loop-cluster $\mathcal{F}_1 \leq \alpha V^{1/3}$\cite{PhysRevE.103.012102,PhysRevE.97.022107}: $P(\mathcal{F}_1\leq V^{\frac{1}{3}})=\langle\mathcal{P}^{(1)}\rangle$,$P(\mathcal{F}_1\leq 2V^{\frac{1}{3}})=\langle\mathcal{P}^{(2)}\rangle$;\par
Moreover, we measure the following quantities in the FK representation:
\begin{enumerate}[label=(\alph*)]
    \item  The sizes of the first- and second-largest FK clusters $\mathcal{C}_1,\mathcal{C}_2$ and their average $C_1=\langle \mathcal{C}_1\rangle, C_2=\langle \mathcal{C}_2\rangle$ the average size of the largest FK cluster conditioned on the origin loop configuration where $\mathcal{F}_1\leq 2V^{1/3}$, denote as $C_1^{\rm P'}$;
    \item The total size of loop clusters in the first- and second-largest FK clusters: $\mathcal{S}_{C_1}=\sum_{\mathcal{F}_i \subset \mathcal{C}_1} \mathcal{F}_i$, $\mathcal{S}_{C_2}=\sum_{\mathcal{F}_i \subset \mathcal{C}_2}\mathcal{F}_i$, and the average of them divided by the total loop cluster size $\mathcal{N}_{\rm v}$ as $n_{\rm f,1}=\frac{\langle\mathcal{S}_{C_1}\rangle}{\langle\mathcal{N}_{\rm v}\rangle}, n_{\rm f,2}=\frac{\langle\mathcal{S}_{C_2}\rangle}{\langle\mathcal{N}_{\rm v}\rangle}$.
\end{enumerate}
\section{Theoretical analysis}
\label{secIII}
\label{theoretical analysis}
The CG Ising model can be exactly solved in its spin representation~\cite{luijten1997interaction}. Hereby, we derive the exact solution of some properties, especially the average number of bond $B$, in the loop representation from the spin representation.\par
The total energy of the CG-Ising model gives
\begin{equation}
    E = -\frac{1}{2V}\sum_{i\neq j}s_js_j=-\frac{1}{2}(Vm^2-1).
    \label{eq:E_spin}
\end{equation}
where $m=(\sum_{i=1}^V s_i)/V$ is the magnetization in the spin representation.
% For the Ising model \eqref{eq:model_spin} in the spin representation with the interaction strength $J_c = 1/V$, we assume that there are $\frac{V}{2}(1+m)$ spins with spin value $+1$ and $\frac{V}{2}(1-m)$ with the magnetic density $m$, such that the total energy gives 
%  \begin{align}
%       \label{eq:E_spin}
%       E &= -\frac{1}{V} [ {2 \choose \frac{V(1+m)}{2}} + {2 \choose \frac{V(1-m)}{2}} - \frac{V(1+m)}{2} \cdot \frac{V(1-m)}{2} ]  \nonumber \\
%         &=-( \frac{1}{2}Vm^2 - \frac{1}{2}). 
%  \end{align}
The probability density function of the magnetization at the critical point is~\cite{luijten1997interaction}
\begin{equation}
    f(m)=\frac{\exp(-\frac{1}{12}Vm^4)}{\int_{-\infty}^{\infty}\exp(-\frac{1}{12}Vz^4)dz},
    \label{eq:distrofm}
\end{equation}
from which we can derive the critical average magnetic density $\langle m^2 \rangle$ as
        \begin{equation}
            \langle m^2 \rangle = 2\sqrt{3} \frac{\Gamma(3/4)}{\Gamma(1/4)} \frac{1}{\sqrt{V}}- \frac{12}{5} \left[ \frac{\Gamma(3/4)}{\Gamma(1/4)} \right]^2 V^{-1} + O(V^{-3/2}),
        \end{equation}
where $\Gamma(\cdot)$ refers to the Gamma function.
Thus, the energy at criticality behaves as
            \begin{equation}
                \langle E \rangle = -\sqrt{3}\frac{\Gamma(3/4)}{\Gamma(1/4)} \sqrt{V}+ \frac{6}{5} \left[ \frac{\Gamma(3/4)
                }{\Gamma(1/4)} \right]^2 +\frac{1}{2} + O(V^{-1/2})\;.  
            \end{equation}
For the loop representation, the partition function can be written as 
        \begin{equation}
            \scrZ= 2^{V} \cosh^{|\mathcal{E}|}{\left(\frac{K}{V}\right)} \sum_{\mathcal{F}\in {\rm even}(G)} \tanh^{|\mathcal{F}|}{\left(\frac{K}{V}\right)}.
        \end{equation}
Here, $|\mathcal{F}|\equiv \scrB(\mathcal{F})$ is the number of bonds on the loop configuration $\mathcal{F}$.
Then the average energy can be calculated as
        \begin{align}
            \langle E \rangle &= - \frac{1}{\scrZ} \frac{\partial \scrZ}{ \partial K } \nonumber \\ 
            &= -\frac{\tanh{\left(\frac{K}{V}\right)}}{V} \left[|\mathcal{E}| + \langle \scrB \rangle \sinh^{-2}{\left(\frac{K}{V}\right)}\right] .    
            \label{eq:E_worm}
        \end{align}
At the critical point $K=1$, since $\tanh(1/V), \sinh(1/V) \approx 1/V$, it follows that $\langle E \rangle=-(\langle\scrB \rangle + \frac{1}{2} - \frac{1}{2V})$. Combining with Eq.~\eqref{eq:E_worm}, we can obtain the leading term of the average value of bond number, 
        \begin{equation}
            \langle \scrB\rangle = \sqrt{3}\frac{\Gamma(3/4)}{\Gamma(1/4)} \sqrt{V} + O(1),
            \label{Nb_analysis}
        \end{equation}
where the amplitude $\sqrt{3}\frac{\Gamma(3/4)}{\Gamma(1/4)} = 0.585414\cdots$.\par
From Eq.~\eqref{eq:E_spin} and Eq.~\eqref{eq:distrofm}, we can also obtain the distribution of the energy on the CG-Ising model as 
\begin{equation}
    f(E)= A_EV^{-\frac{1}{4}}\exp\left[-\frac{(1-2E)^2}{12V}\right]\sqrt{\frac{1}{(1-2E)}},
    \label{eq:Edist}
\end{equation}
with the normalized factor $A_E = \frac{3^{-1/4}}{\sqrt{2}}\Gamma^{-1}(\frac{5}{4})$. Here, we assume that the probability distribution of bond number is also equivalent to the one of the total energy, which is similar to the relation of the average. By replacing $E$ with $-(\scrB+1/2)$ in Eq.~\eqref{eq:Edist} we conjecture the distribution of bond number is
\begin{equation}
    f(\scrB)= A_BV^{-\frac{1}{4}}\exp\left(-\frac{\scrB^2}{3V}\right)\sqrt{\frac{1}{\scrB}},
    \label{dist_NB}
\end{equation}
where the normalized factor $A_B = A_E/\sqrt{2}= 0.419149\cdots$. 
% Thus, the probability of the vacant configuration is naturally exhibited as $b=0$:
% \begin{equation} 
%     P^{\rm v} = A_BV^{-1/4}
% \end{equation}
\begin{figure}[h]
    \centering
    \includegraphics[width=0.5\textwidth]{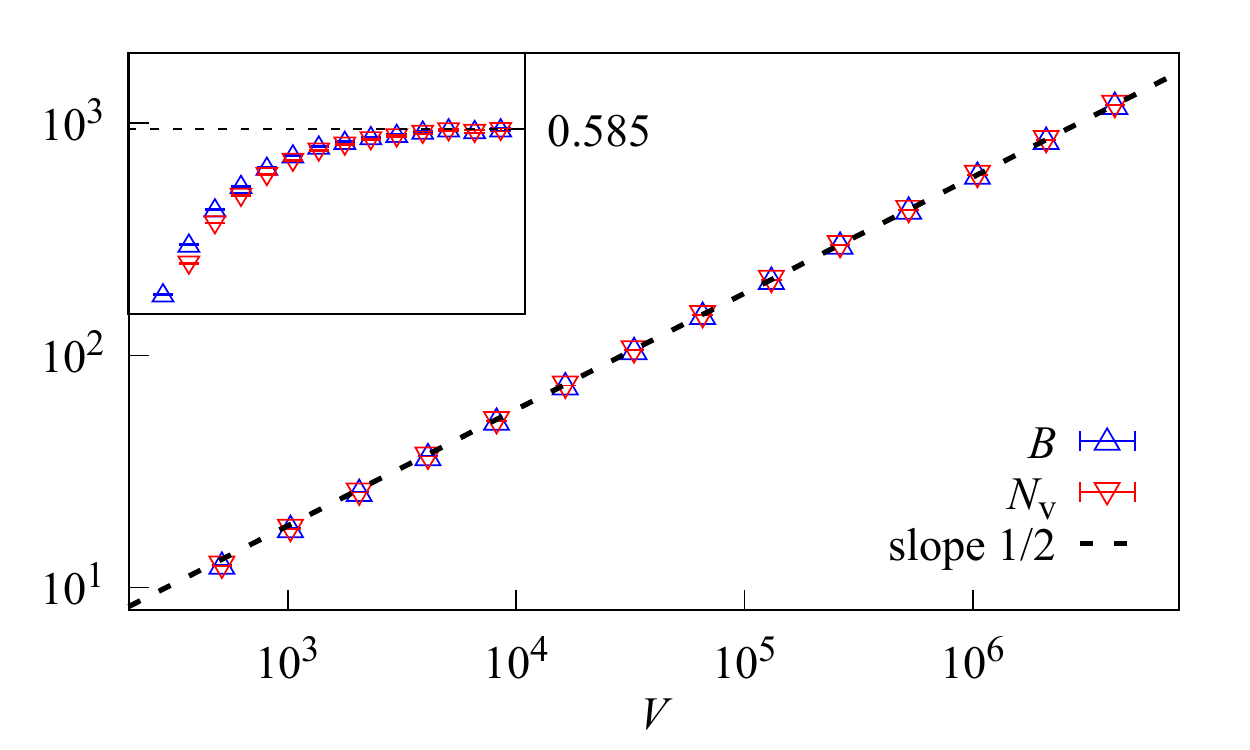}
    \caption{The log-log plot of the bond number $B$ and total number of vertices in the loop clusters $N_{\rm v}$ versus the system volume $V$. The inset displays the rescaled terms $N_{\rm v}/\sqrt{V},B/\sqrt{V}$ versus $V$.}
    \label{fg:NB}
\end{figure}
\section{Numerical results}
\label{secIV}
\subsection{Scaling behaviors of geometric quantities}
% {Scaling behaviors of important quantities in the loop representation  \blue{(Critical scaling behaviors of geometric quantities)}}

% \blue{We perform least-square fits on the FSS (FSS) data. As a precaution against correction-to-scaling terms that we missed including in the fitting ansatz, we impose a lower cutoff $V \ge \Vm$ on the data points admitted in the fit and systematically study the effect on the residuals $\chi^2$ value by increasing $\Vm$. In general, the preferred fit for any given ansatz corresponds to the smallest $\Vm$ for which the goodness of the fit is reasonable and for which subsequent increases in $\Vm$ do not cause the $\chi^2$ value to drop by vastly more than one unit per degree of freedom. In practice, by “reasonable” we mean that $\chi^2/\rm{DF} \approx 1$, where DF is the number of degrees of freedom. The systematic error is estimated by comparing estimates from various sensible fitting ansatz.}  \par  

In this section, we study the scaling behaviors of some geometric quantities such as the number of bonds $B$ and the number of clusters $N_{\rm k}$, and the sizes of the first- and second-largest clusters $F_1,F_2$ in the loop representation. We perform least-square fits to our data. As a precaution against correction-to-scaling terms that we missed including in the fitting ansatz, we impose a lower cutoff $V \ge \Vm$ on the data points admitted in the fit and systematically study the effect on the residuals $\chi^2$ value by increasing $\Vm$. In general, the preferred fit for any given ansatz corresponds to the smallest $\Vm$ for which the goodness of the fit is reasonable and for which subsequent increases in $\Vm$ do not cause the $\chi^2$ value to drop by vastly more than one unit per degree of freedom. In practice, by “reasonable” we mean that $\chi^2/\rm{DF} \approx 1$, where DF is the number of degrees of freedom. The systematic error is estimated by comparing estimates from various sensible fitting ansatz.
% using the finite size scaling(FSS) method. These analyses allow us to derive the fractal dimension $d_F$, observe the critical phenomenon in the loop representation and verify whether it is different from the FK representation.\par
\begin{table}[]
\setlength\tabcolsep{2.0pt}
\begin{tabular}{lllllll}
\hline\hline
$\mathcal{O}$ & ~~~~$y_\scrO$       & ~~~$a_0$      & ~~~$b_1$      & ~~~$y_1$    & $\chi^2/DF$ & $V_{\rm min}$ \\ \hline 
%$\mathcal{O}$ & $y_0$       & $a_0$      & $b_1$      & $y_1$    & $\chi^2/DF$ & $V_{\rm min}$ \\ \hline 
              & 0.499~9(2)   & 0.586(2)   & -0.68(7)   & -0.49(2) & 7.1/9     & $2^{10}$      \\
              & 0.499~8(3)   & 0.587(2)   & -0.65(12)  & -0.48(3) & 7.0/8     & $2^{11}$      \\
              & 0.500~1(3)   & 0.584(2)   & -1.1(4)    & -0.55(6) & 4.7/7     & $2^{12}$      \\
$B$           & 0.499~98(6)  & 0.585~6(5)  & -0.725(8)  & -1/2     & 7.5/10    & $2^{10}$      \\
              & 0.499~96(8)  & 0.585~7(6)  & -0.728(13) & -1/2     & 7.4/9     & $2^{11}$      \\
              & 1/2          &0.585~3(1)   &-0.73(4)    &-0.501(7)  &7.6/10     &$2^{10}$ \\
              & 1/2          &0.585~4(1)   &-0.74(6)    &-0.503(11)  &7.4/9    &$2^{11}$ \\\hline
              & 0.499~8(2)   & 0.457(1)   & 0.44(9)    & -0.54(4) & 7.2/9     & $2^{10}$      \\
              & 0.499~8(2)   & 0.457(2)   & 0.5(2)     & -0.55(6) & 7.1/8     & $2^{11}$      \\
$F_1$         & 0.500~2(4)   & 0.455(3)   & 0.2(1)     & -0.4(1)  & 5.4/7     & $2^{12}$      \\
              & 0.500~09(7)  & 0.455~5(3)  & 0.361(7)   & -1/2     & 8.4/10    & $2^{10}$      \\
              & 0.500~06(8)  & 0.455~7(4)  & 0.356(10)  & -1/2     & 7.9/9     & $2^{11}$      \\ \hline
            & 0.499~6(2)         & 0.092~5(2)         & 1.13(7)   & -0.68(1)  & 9.1/9 & $2^{10}$ \\ 
            & 0.499~6(2)         & 0.092~6(3)         & 1.12(13)  & -0.68(2) & 9.1/8 & $2^{11}$ \\
$F_2$       & 0.499~8(3)         & 0.092~3(4)         & 0.9(2)  & -0.66(3) & 8.5/7 & $2^{12}$ \\
            & 0.499~81(9)  & 0.092~3(1)  & 1.023(8)  & -2/3      & 9.7/9 & $2^{11}$ \\
            & 0.499~7(1)   & 0.092~4(1) & 1.01(3) & -2/3      & 8.6/8 & $2^{12}$ \\ 
\hline\hline
\end{tabular}
%\caption{Fits of $F_1,F_2,B$ in the loop representation}
  \caption{The fitting results of the bond number $B$, the first- and second-largest clusters $\Clone$, $\Cltwo$. We conjecture all of them have the same scaling behavior $\Clone, \Cltwo, B \sim \sqrt{V}$, which suggests there is no two-length scaling behavior in the loop representation. }
\label{table:Fit1}
\end{table}
\par 

% Generally, we use the following ansatz to describe the FSS behavior of quantity $\mathcal{O}$ and perform a least-square fit with our MC data to estimate its critical exponent $y_\mathcal{O}$:
We first consider the number of bonds $B$. In Fig.~\ref{fg:NB}, we plot $B$ versus the system volume $V$ in log-log scale, and the dashed line with slope $1/2$ suggests $B \sim \sqrt{V}$. Meanwhile, the inset plots $B/\sqrt{V}$ showing that its amplitude tends to 0.585. These results are consistent with our theoretical analysis in Eq.~\eqref{Nb_analysis}.\par

To extract the scaling behaviors of $B$, we perform the least-square fits via the general ansatz:
\begin{equation}
\mathcal{O}=V^{y_\mathcal{O}}(a_0+b_1V^{y_1}+b_2V^{y_2})+c,
\label{equtaion: est}
\end{equation}
where $\mathcal{O}$ corresponds to the quantities measured, such as $B$, and $y_{\mathcal{O}}$ corresponds to the dominant scaling exponent as $y_{\textsc b}$ for $B$. For $B$, we first leave all parameters free, which gives unstable results. We then fix $b_2=c=0$ and leave $y_{\textsc b}$, $a_0$, $b_1$ and $y_1$ free, and it gives reasonable estimate $y_{\textsc b} = 0.499\,8(3)$ and $y_1 =-0.48(3)$ for $\Vm = 2^{11}$. We then try to fit by fixing $y_1=-1/2$, as predicted in Eq.~\eqref{Nb_analysis}, and the fitting gives a reasonable estimate $y_{\textsc b}= 0.499\,96(8)$. More trials have been tried, like fixing $b_1=b_2=0$ and leaving $y_{\textsc b}$, $a_0$ and $c$ free, which gives consistent results. Including the systematic errors by comparing various reasonable results, we finally obtain the estimates $y_{\textsc b} = 0.500\,0(8)$ and $a_0 =0.585(6)$, both of which are consistent with Eq.~\eqref{Nb_analysis}. The fitting details are summarized in Table~\ref{table:Fit1}. \par
\begin{figure}[h]
    \centering
    \includegraphics[width=0.98\linewidth]{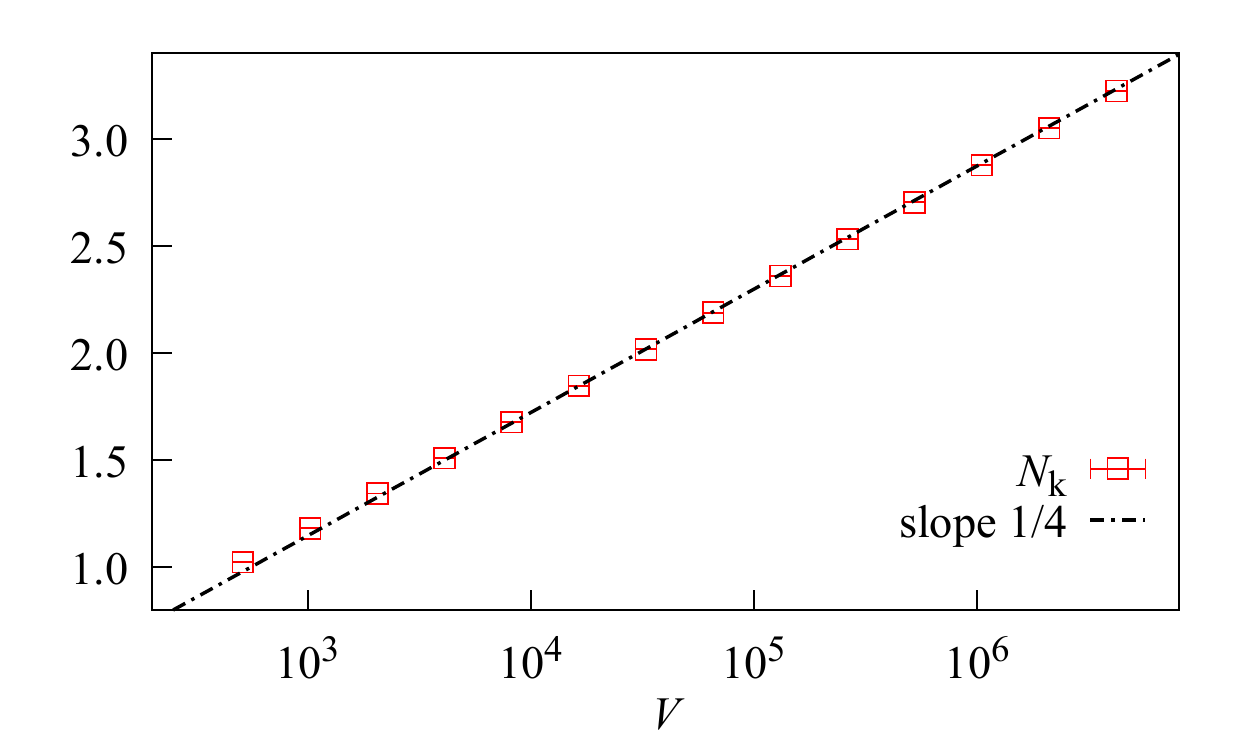}
%    \caption{semi-log plot of $N_{\rm k}$ versus $V$. The inset shows $N_{\rm k}/\ln{V},N_c/\ln{V}$ versus $V$.}
     \caption{The semi-log plot of the cluster number $N_{\rm k}$ versus $V$, which suggests it scales as $N_{\rm k} \sim \frac{1}{4}\ln V$. }
    \label{fg:Nc}
\end{figure}
For a given loop configuration, a loop cluster is defined as a set of vertices which are connected together by loop bonds. We next study the geometric properties of these loop clusters. In the graph theory, we have the Euler formula $N_{\rm v} - B = N_c -N_{\rm k}$  with the number of cycles $N_c$ and the number of clusters $N_{\rm k}$. It inspires us to observe whether the Eulerian clusters in the loop representation are uni-cyclic or multi-cyclic by evaluating $N_{\rm v}$, since for the uni-cyclic graph, $N_c$ equals to $N_{\rm k}$. Figure~\ref{fg:NB} presents the FSS behavior of $N_{\rm v}$ and $N_{\rm v}/\sqrt{V}$ in the inset. It suggests that the value of $N_{\rm v}$ is numerically consistent with $B$ as $V$ is large enough. Therefore, we can argue that  $N_c = N_{\rm k}$ in thermodynamic limit, which means that all the loop clusters are asymptotically uni-cyclic in the thermodynamic limit.\par
\begin{table}[]
\begin{tabular}{lllllll}
\hline
$\mathcal{O}$ & $a$       & $b_1$   & $y_1$    & $c$       & $\chi^2/DF$ & $V_{\rm min}$ \\ \hline
         & 0.249~8(4) & 1.1(2)  & -0.49(4) & -0.587(6) & 4.7/9     & $2^{10}$      \\
  $N_c$            & 0.249~6(4) & 1.4(6)  & -0.53(8) & -0.584(7) & 4.4/8     & $2^{11}$      \\
              & 0.249~7(1) & 1.17(3) & -1/2     & -0.586(2) & 4.8/10    & $2^{10}$      \\
              & 0.249~8(2) & 1.19(4) & -1/2     & -0.587(2) & 4.6/9     & $2^{11}$      \\ \hline
\end{tabular}
%\caption{Fits of $N_c$ in the loop representation}
\caption{The fitting result of the cluster number $N_{\rm k}$, which scales as $N_{\rm k} \sim \ln V$ with the coefficient consistent with 1/4.}
\label{table:Fit2}
\end{table}
Besides, we also study the scaling behavior of the number of loop clusters. As shown in Fig.~\ref{fg:Nc}, our data of $N_{\rm k}$ collapse onto the dashed line with slope $1/4$ in semi-log scale, indicating that $N_{\rm k} \sim \frac{1}{4}\ln V$. We can fit the data of $N_{\rm k}$ to the ansatz,
\begin{equation}
  N_\mathcal{O}= a\ln{V}  + b_1V^{y_1}+b_2V^{y_2} + c.
    \label{eq:fit2}
\end{equation}
We first leave all parameters free, but there is no stable fit. Then by fixing $b_2=0$, we obtain stable fits, with details shown in Table~\ref{table:Fit2}. We estimate $a=0.249~7(5)$ which leads to a conjecture $N_{\rm k} \asymp \frac{1}{4}\ln{V}$. \par
\begin{figure}[h]
    \centering
    \includegraphics[width=0.5\textwidth]{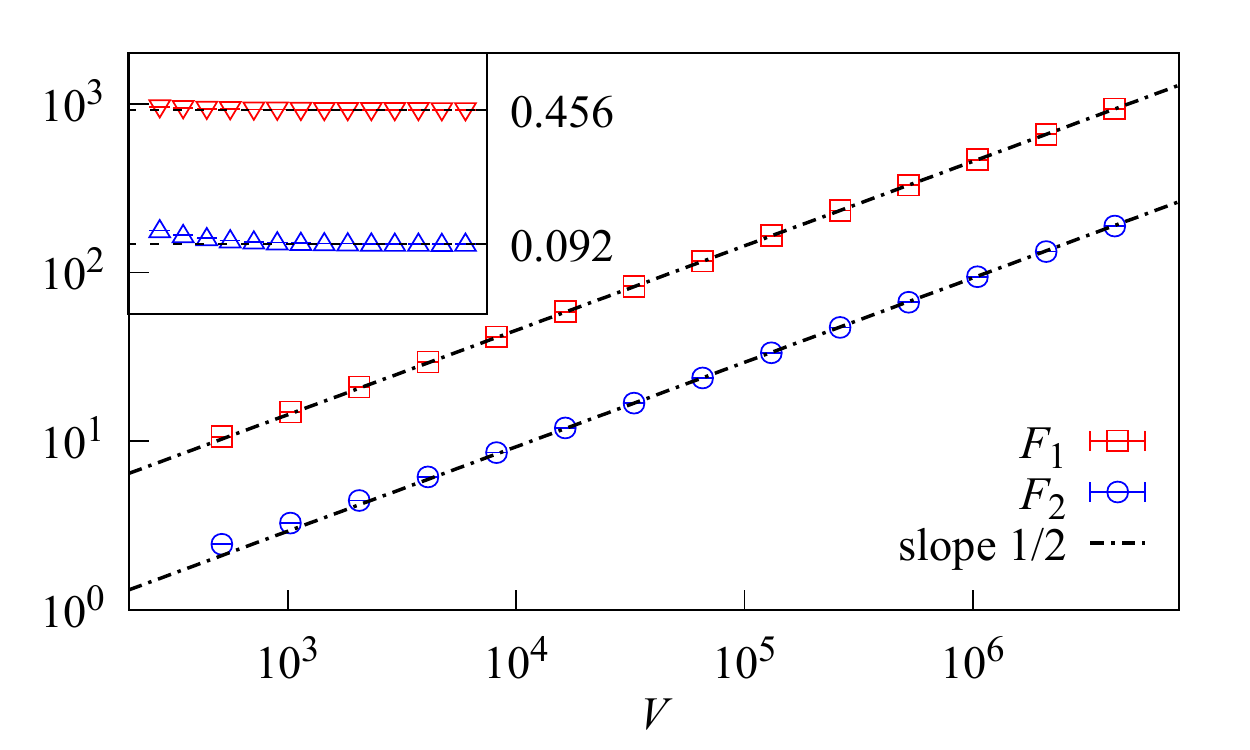}
    \caption{The log-log plot of the size of the largest cluster $F_1$, the second-largest cluster $F_2$ versus the system volume $V$. Our data suggest $F_1,F_2 \sim \sqrt{V}$. The inset displays the rescaled terms $F_1/\sqrt{V},F_2/\sqrt{V}$ versus $V$. }
    \label{fg:Ln}
\end{figure}

We then consider the sizes of the largest cluster $F_1$ and second-largest cluster $F_2$. As Fig.~\ref{fg:Ln} shows, we plot $F_1$ and $F_2$ in the log-log plot, and the slope $1/2$ indicates both of them have the same scaling behavior $F_1, F_2 \sim \sqrt{V}$. In other words, no two-length scaling behavior has been observed, which is different from the observation of two largest clusters in the FK Ising model on the CG~\cite{luczak2006phase,PhysRevE.103.012102}.
\par 
We also perform the least-square fit via \eqref{equtaion: est} for $F_1$, $F_2$ as $y_{\scrO}$ corresponds to the volume fractal dimensions $d_{\rm f1}$ and $d_{\rm f2}$, respectively. The fitting results through different trails are reported in Table~\ref{table:Fit1}. We  obtain the estimates $d_{\rm f1} = 0.500\,0(9)$ and $a_0 =0.456(6)$ for $F_1$ while $d_{\rm f2} = 0.499\,7(6)$ and $a_0 = 0.092\,4(8)$ for $F_2$. We found both $d_{\rm f1}$ and $d_{\rm f2}$ are consistent with $1/2$, and the amplitude $a_0$ of $F_2$ is much smaller than that of $F_1$.

\par 

%  Besides, we can also study the scaling behavior of the number of loop cluster straightforwardly. As we can see in Fig.~\ref{fg:Nc}, the dash line with slope $1/4$ content to our data of $N_{\rm k}$ on semi-log coordinate indicates $N_{\rm k}$ scale as $\frac{1}{4}\ln{V})$. Analogous to the fitting function of the bridge-free cluster of percolation on CG in Ref.~\cite{PhysRevE.97.022107}, We can use this ansatz to describe the scaling behavior and perform a least-square fit:
% \begin{equation}
%   N_\mathcal{O}= a\ln{V}  + b_1V^{y_1}+b_2V^{y_2} + c.
%     \label{eq:fit2}
% \end{equation}
% We set $b_2=0$ for a stable fit result and the fit result is shown Table~\ref{table:Fit2}. We estimate $a=0.249~7(5)$, so we can conjecture that $N_{\rm k} \sim \frac{1}{4}\ln{V}$. In the inset of Fig.~\ref{fg:Nc}, we plot $N_{\rm k}/\ln V$ and $N_c/\ln V$ versus system volume $V$, which suggests both of them converge to the same value and directly shows that all loop clusters are uni-cyclic.

\subsection{Probability distribution of geometric quantities}

\begin{figure}[ht]
    \centering
    \includegraphics[width=0.98\linewidth]{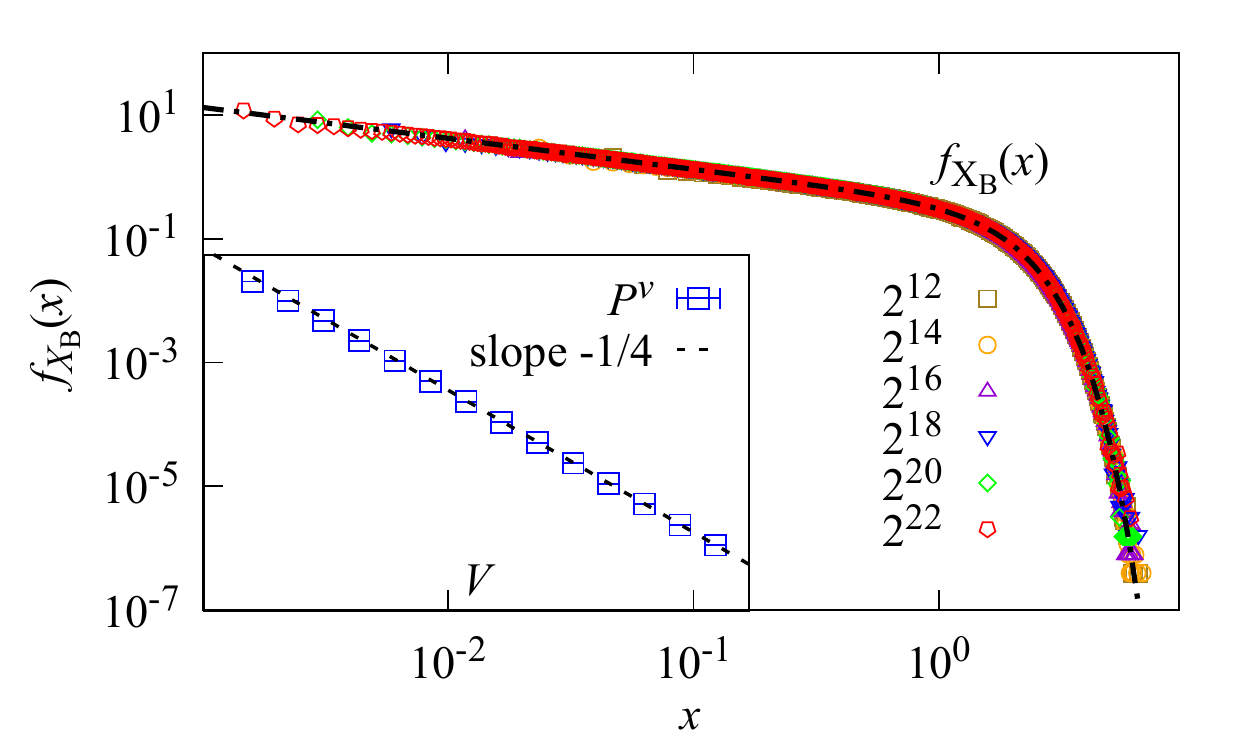}
    \caption{Log-log plot of probability distribution of the number of bonds $\scrB$ on the CG in the loop representation.  Here $f_{X_B}(x)$ is the probability density function of $X_B = \scrB/\sqrt{V}$. The solid black line referring to $f_0(x)=A_B\exp(-x^2/3)x^{-1/2}$ verifies that the numerical result is consistent to the theoretical analysis based on the spin representation. The inset shows the probability of vacant graph $P^{\rm v}$ versus $V$ with log-log plot, which implies $P^{\rm v}$ exhibits a power-law decay as $V^{-1/4}$.}
    \label{fig:dist_NB}
\end{figure}

Firstly, we investigate the probability distribution of the number of bonds $\scrB$. Denote $f(\scrB,V)$ the probability density function (PDF) of $\scrB$ sampled in our simulations. Since $B \sim \sqrt{V}$, we define $X_B=\scrB/\sqrt{V}$ and $f_{X_B}(x)$ the PDF of $X_B$. Then it follows that
\begin{equation}
    f(\scrB,V)d\scrB = f_{X_B}(x)dx,
\end{equation}
where $\sqrt{V} dx= d\scrB$, and thus $f_{X_B}(x)=\sqrt{V}f(\scrB,V)$. 
From Eq.~\eqref{dist_NB}, one obtains $f_{X_B}(x) = A_B\exp(-x^2/3)x^{-1/2}$ with  $A_B = \frac{3^{-1/4}}{2}\Gamma^{-1}(\frac{5}{4})$. Figure~\ref{fig:dist_NB} presents the distribution of $X_B$, and the dashed curve displays $f_{X_B}(x)$.
%The numerical result has been shown in Fig.~\ref{fig:dist_NB}, where the solid black line display the theoretical distribution function of $\mathcal{N}_B$ derived from Eq.~\eqref{dist_NB} as $f_0(x)= A_B\exp(-x^2/3)x^{-1/2}$.
It is obvious that our numerical result is consistent with the theoretical analysis. Besides, we found out the probability of the vacant graph (no occupied bond) $P^{\rm v}$ obeys a power-law decay as $V^{-1/4}$, as suggested by the log-log plot of $P^{\rm v}$ versus $V$ in the inset of Fig.~\ref{fig:dist_NB} and Eq.~\eqref{dist_NB}. We perform a least-square fit to the ansatz Eq.~\eqref{equtaion: est} and estimate the power-law exponent of $P^{\rm v}$ as $y_\scrO=-0.249(1)$ and the coefficient $a = 1.56(3)$.\par
\begin{figure}[ht]
    \centering
    \includegraphics[width=0.98\linewidth]{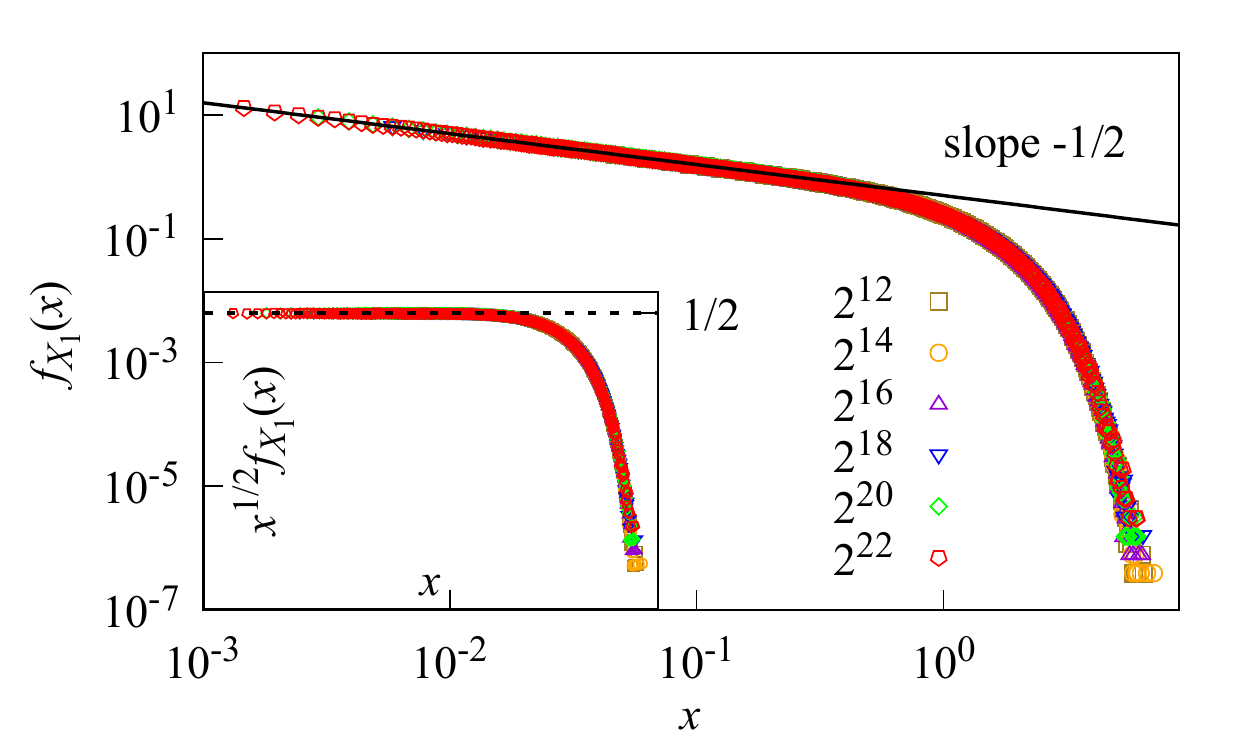}
    \caption{Log-log plot of probability distribution of the largest cluster $\mathcal{F}_1$ on the CG in loop representation, where $f_{X_1}(x)$ is the probability density function of $X_1 = \mathcal{F}_1/\sqrt{V}$. The inset shows the log-log plot of $f_{X_1}(x) x^{1/2}$ versus the rescaled variable $x$.}
    \label{fig:L1ds}
\end{figure}

% In Ref.\cite{PhysRevE.103.012102}, the emergence of two configuration sectors in the FK representation on the CG was discovered from the anomalous scaling behavior of the probability distribution of the largest cluster. Therefore, to discover whether the loop representation exhibits a similar scaling behavior, we study $f_{\mathcal{F}_1}(s,V)$, the probability density function of $\mathcal{F}_1$.\par
Then we study $f_{\mathcal{F}_1}(s,V)$, the PDF of $\mathcal{F}_1$.
Since its mean scales as $\sqrt{V}$, we first study the distribution of $X_1 := \scrF_1/\sqrt{V}$.
Figure~\ref{fig:L1ds} presents the PDF of $f_{X_1}(x)$ versus $x$ in the log-log scale.
The excellent data collapse suggest that $f_{X_1}(x)$ follows a power-law distribution
\begin{equation}
    f_{X_1}(x) \asymp x^{-1/2}\Tilde{f}(x) \;,
    \label{dist_F1}
\end{equation}
with $\Tilde{f}(x\to 0) \approx 1/2$ when $x$ is small and $\Tilde{f}(x)$ decays quickly to zero when $x$ is large, as indicated from the inset of Figure~\ref{fig:L1ds}.

\begin{figure}[ht]
    \centering
    \includegraphics[width=0.98\linewidth]{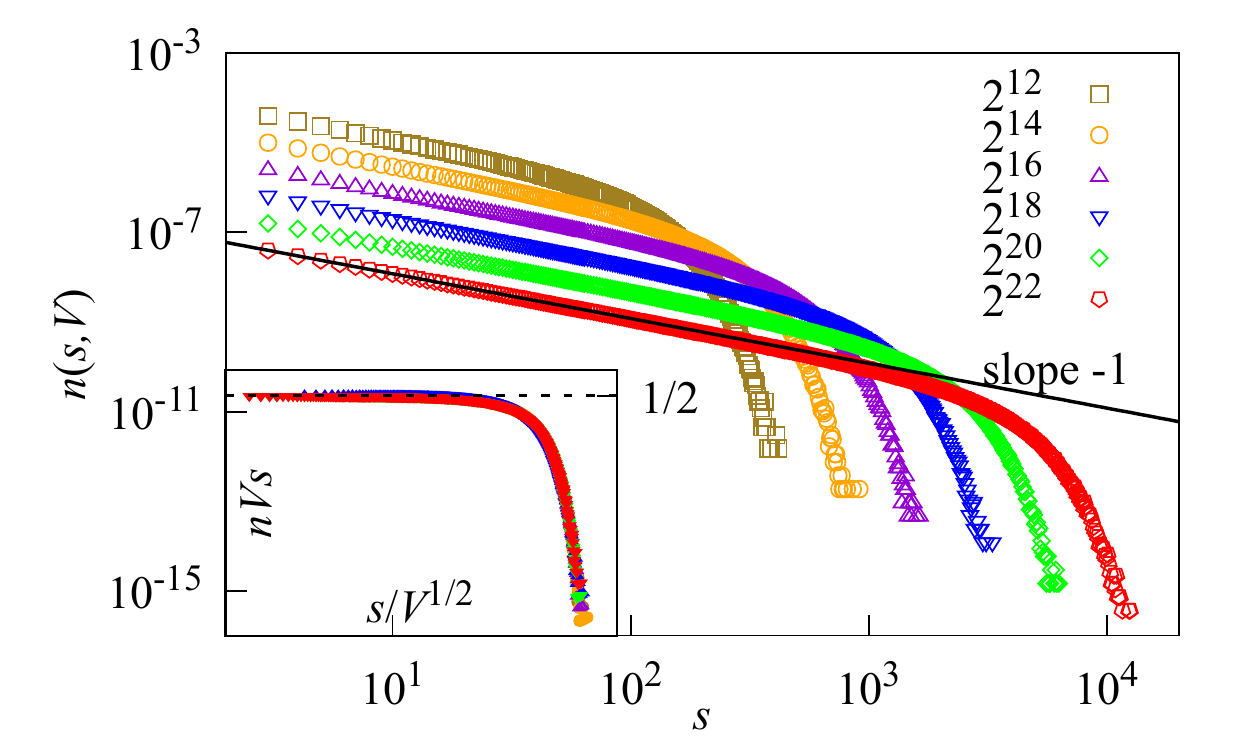}
    \caption{Cluster-number density of the loop representation in the log-log scale. The inset shows the plot of $n(s,V)Vs$ versus $s/\sqrt{V}$, which implies that the scaling function is consistent with $1/2$ when $s/\sqrt{V} \ll 1$.}
    \label{fig:nsdv}
\end{figure}

Meanwhile, we also study the cluster-number density of the loop representation $n(s,V)$. 
% The standard FSS\cite{PhysRevE.71.026129} predicts that on the CG the cluster-size distribution follows the form $n(s,V)\sim s^{-\tau}n(s/V^{d_F})$, with the Fisher exponent $\tau = 1 + 1/d_F$ related to the fractal dimension. In the previous section, it's known to us that the fractal dimension in the loop representation on the CG is $d_F = 1/2$, so the standard Fisher exponent should be $\tau = 3$.\par
Our results of $n(s,V)$ on the CG, shown in Fig.~\ref{fig:nsdv}, indicate that it follows the form $n(s,V)\sim s^{-\tau}\tilde{n}(s/V^{d_F})$ with a modified Fisher exponent $\tau = 1$. More specifically, we can conjecture that the distribution obeys 
\begin{equation}
    n(s,V) \asymp n_0s^{-\tau}V^{-h}\tilde{n}(s/V^{d_F}) \;,
    \label{equation:dis_n}
\end{equation}
where $h \geq 0$ is the scaling exponent, $\tilde{n}(x)$ is scaling function which is approximately 1 when $x$ is small. This leads to the number of loop clusters as the integral of $n(s,V)$ from $1$ to the largest loop cluster
\begin{equation}
 N_{\rm k}= V \int_{1}^{F_1}n(s,V)ds.
\end{equation}
Our previous results suggest $F_1\sim a_0\sqrt{V}$, so it follows that
\begin{equation}
    N_{\rm k} \asymp \left\{
             \begin{array}{lr}
             \frac{n_0}{2}V^{1-h}(\ln{V}+\ln{a_0})\ \ &\rm{if}\ \ \tau =1,   \\
             \frac{n_0}{1 -\tau}a_0^{1-\tau}V^{1+\frac{1}{2}(1-\tau)-h}\ \ &\rm{if}\ \ \tau \neq 1.
             \end{array}
\right.
\end{equation}

In the previous section, we know $N_{\rm k} \asymp \frac{\ln{V}}{4}$. Therefore, we obtain  $n_0 = \frac{1}{2},h=1,\tau= 1$. The inset of Fig.~\ref{fig:nsdv} confirms our conjecture, including $n_0 = \frac{1}{2}$.\par

% In Ref.\cite{PhysRevE.103.012102}, the emergence of two configuration sectors in the FK representation on the CG was discovered from the anomalous scaling behavior of the probability distribution of the largest cluster. Therefore, to discover whether the loop representation exhibits a similar scaling behavior, we study $f_{\mathcal{F}_1}(s)$, the probability density function of $\mathcal{F}_1$.\par
% In our simulation, We observe an interesting fact that the number of configuration with $\mathcal{F}_1 = 1$, which means it is empty with bonds, is considerable. It is a starting point for us to discover the scaling behavior of $f_{\mathcal{F}_1}(s)$. Figure.~\ref{fig:Vac} shows log-log plot of the probability of the empty graph $P^v$ versus $V$ which indicates that it obeys a power-law decay as the system volume goes to infinity.We perform a fit through \eqref{equtaion: est} and derive the damping exponent of $P^v$ as $y_0=-0.249(1)$ and the coefficient $a = 1.56(3)$, from which we conjecture that $P^v \sim V^{-\frac{1}{4}}$.
Therefore, in contrast to the FK representation, the scaling behaviors of $f_{{\mathcal F}_1}(s,V)$ and $n(s,V)$ both show that there is only one scaling sector and one length scale in the loop representation.\par
% We also study the probability distribution of the number of bonds in the loop representation $f_{\mathcal{N}_B}(b,V)$, which can also be rescaled as $f_{X_B}(x)$ with $X_B = \frac{\mathcal{N}_B}{\sqrt{V}}$. The lower panel of Fig.~\ref{fig:L1ds} shows the log-log plot of the rescaled function $f_{X_B}(x)$ versus the rescaled variable $x$. The inset of the figure suggests that $f_{X_B}(x)$ is similar to $f_{X_1}(x)$, also decay as $x^{-1/2}$. This result is not surprising, for the reason that the density of loop cluster is tend to zero and the clusters are mostly uni-cyclic.\par

\subsection{Insights for the anomalous FSS behaviors in the random-cluster representation}         
\begin{figure}
    \centering
    \includegraphics[width=0.98\linewidth]{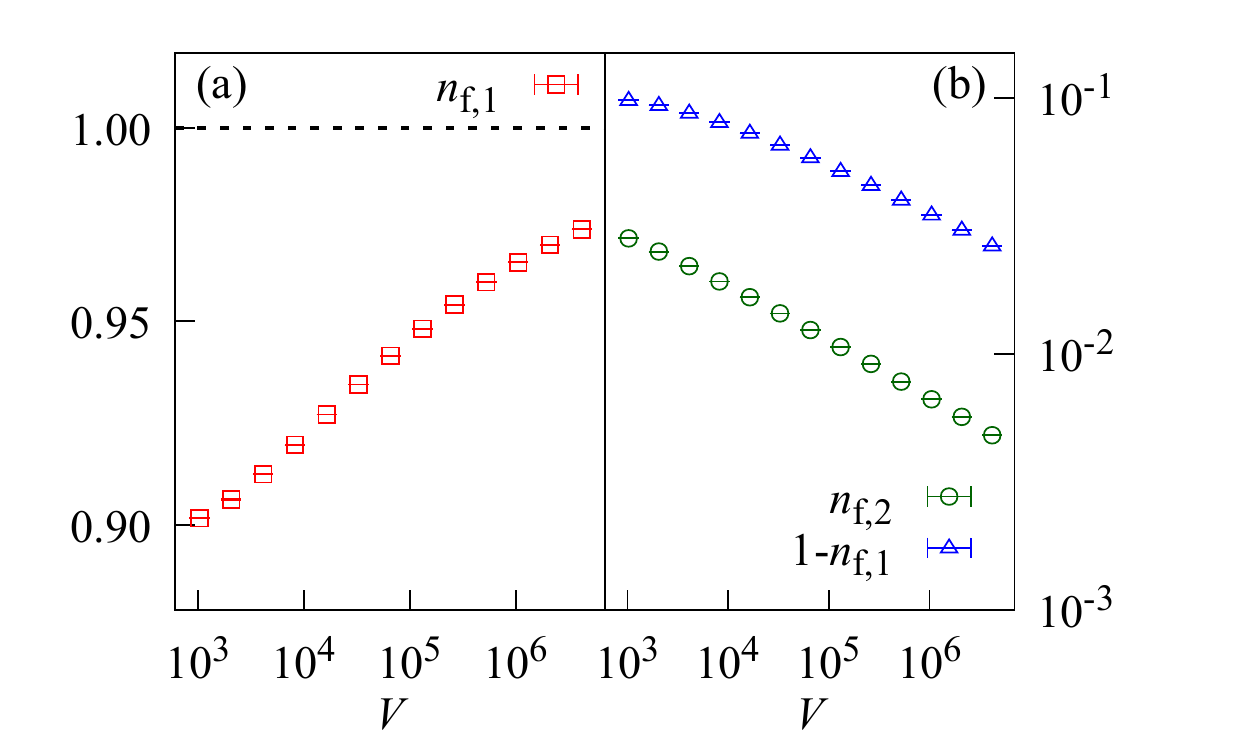}
    \caption{Panel (a) plots the relative loop vertices in the largest FK cluster  $n_{\rm f,1}$ versus $V$. Panel (b) shows the log-log plot of the relative loop vertices in the second-largest FK cluster   $n_{\rm f,2}$ and the one out of the largest FK cluster $1-n_{\rm f,1}$ versus $V$. It implies that all the loop clusters belong to the largest FK cluster as $V\to \infty$.}
    \label{fig:twolengthscale}
\end{figure}
As discovered in above sections, the loop bond density $\frac{B}{V} \sim \frac{1}{\sqrt{V}}$, so the loop configuration is vacant in the thermodynamic limit. Moreover, the probability of adding bonds through the LC algorithm is asymptotically the same as the critical percolation threshold $1/V$, such that the transformation from the loop representation to the FK representation is almost the process of critical percolation. 
%the loop representation does not exhibit the anomalous critical phenomenon of two scaling sectors, two length scales or two scaling windows observed in the FK representation\cite{PhysRevE.103.012102,luczak2006phase}. %However, the LC joint model reveals that the transformation from the loop representation to the RC representation is a process of percolation, which is consistent with our result of the vanishing density of bond in the loop representation. 
In this section, we will demonstrate how LC algorithm can provide an intuitive understanding to the rich critical phenomenon in the FK representation~\cite{Fang_2022,Fang_2022_2}.\par
Firstly, in the FK representation, the largest and second-largest clusters exhibit distinct scaling behaviors: $C_1 \sim V^{3/4}$ and $C_2 \sim \sqrt{V}\log{V}$. However, as Fig.~\ref{fg:Ln} shows, the first- and second-largest clusters in the loop representation both scale as $\sqrt{V}$. One would wonder what happens in the percolation process of the LC algorithm.
We record the relative mass of the loop clusters belonging to the first- and second-largest FK clusters after the representation transformation denoted as $n_{\rm f,1},n_{\rm f,2}$.
As shown in Fig.~\ref{fig:twolengthscale}(a), the relative loop vertices in $\mathcal{C}_1$ increases to $1$ as the system volume becomes larger. In contrast, Fig.~\ref{fig:twolengthscale}(b) shows that the relative loop vertices in $\mathcal{C}_2$ and out of $\mathcal{C}_1$ exhibit a power-law decay to zero as $V$ increases. Furthermore, we perform a least-square fit to Eq.~\eqref{equtaion: est} for $1-n_{\rm f,1}$, and we obtain the decaying exponent $y_\textsc{d}=0.225(6)$. These evidences suggest that in the thermodynamic limit, all loop clusters belong to the largest FK cluster $C_1$ after the percolation process while cycles in other FK clusters are newly generated in the process of percolation.
%This naturally explains the different critical scaling behaviors between the two graphical representation.
\par
Secondly, two sectors are observed in the configuration space of the FK representation: the percolation sector $S^{\rm P}$ with its size of the largest cluster $\mathcal{C}_1\leq O(V^{2/3})$ and the Ising sector $S^{\rm I}$ for otherwise. The percolation sector vanishes with the rate $V^{-1/12}$ and the largest cluster in this sector scales as $C_1^{\rm P}\sim V^{2/3}$. We are trying to find out what the percolation sector corresponds to in the loop configuration space.\par
\begin{figure}[ht]
    \centering
    \includegraphics[width=0.98\linewidth]{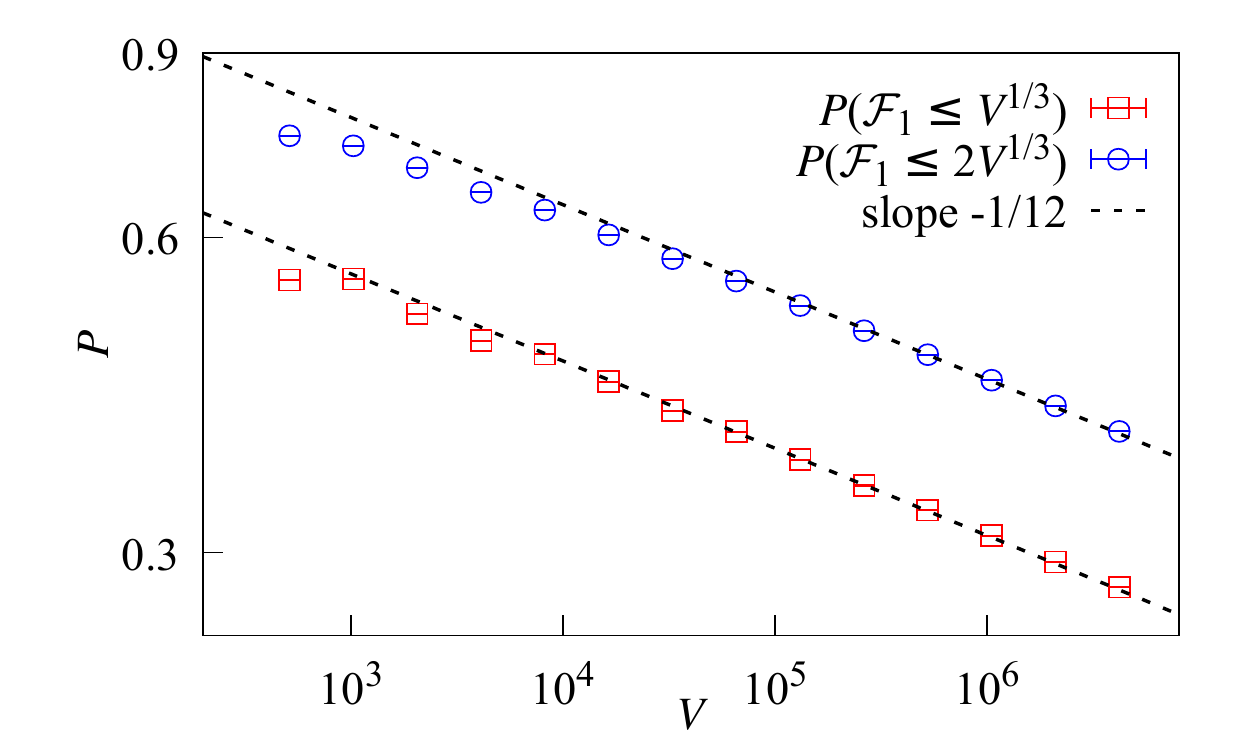}
    \caption{Log-log plot of the probability that the loop configuration in the conjectured percolation sector $P(\mathcal{F}_1 \leq \alpha V^{1/3})$ with $\alpha = 1$ or $2$. The straight dashed black lines with slope $-1/12$ are to guide the eye.}
    \label{fig:PerS}
\end{figure}

In graph theory, a bridge is a bond whose deletion would break a cluster into two.
The configuration with all bridges deleted is called bridge-free configuration. The clusters in the loop representation are all bridge-free clusters. From Ref.~\cite{PhysRevE.97.022107}, we know that the volume fractal dimension of the bridge-free cluster in the CG-percolation model is $d_B=\frac{1}{3}$, so we conjecture that the corresponding percolation sector in the loop representation $S^{\rm P}_l$ consists of the configurations whose $\mathcal{F}_1 \leq \alpha V^{\frac{1}{3}}$ with some constant $\alpha$. The probability of $S^{\rm P}_l$ can be derived from the probability distribution of the largest loop cluster $f_{\mathcal{F}_1}(s,V)$\eqref{dist_F1} :
\begin{align}
    P(S^{\rm P}_l)&=P(\mathcal{F}_1\leq \alpha V^{1/3}) \approx 2\alpha^{\frac{1}{2}}\tilde{f}(0)V^{-\frac{1}{12}}\;,
    \label{perS}
\end{align}
which is perfectly consistent with the probability of the percolation sector in the FK representation as $P(S^{\rm P})\sim V^{-1/12}$. The numerical result of $P(\mathcal{F}_1\leq \alpha V^{1/3})$ with $\alpha=1,2$ versus $V$ is shown in Fig.~\ref{fig:PerS}. We perform a least-square fit to Eq.~\eqref{equtaion: est} with our $P(S^{\rm P}_l)$ data and obtain $y_\mathcal{O} = -0.081(2)$ for $\alpha=1$, which is consistent with $-\frac{1}{12}$. By fixing $y_\scrO= -\frac{1}{12}$, we estimate the coefficient $2\alpha^{\frac{1}{2}}\tilde{f}(0)=0.986(4)$ for $\alpha=1$ and $2\alpha^{\frac{1}{2}}\tilde{f}(0)= 1.388(4)$ for $\alpha=2$. It then follows that $\tilde{f}(0) = 0.48(2)$ which is consistent with our conjecture $1/2$.
%Meanwhile, we can conjecture the coefficient $f_0$ of $f_{\mathcal{F}_1}(s,V)$ from Eq.\eqref{perS} through our fitting result. To eliminate the finite-size correction, we have
%\begin{equation*}
%    f_0 = \frac{a_{(2)}-a_{(1)}}{2(\sqrt{2}-1)}\approx 0.48(2),
%\end{equation*}
%which is consistent with $\frac{1}{2}$. Therefore, we conjecture $f_0=\frac{1}{2}$.\par
\begin{figure}[ht]
    \centering
    \includegraphics[width=0.98\linewidth]{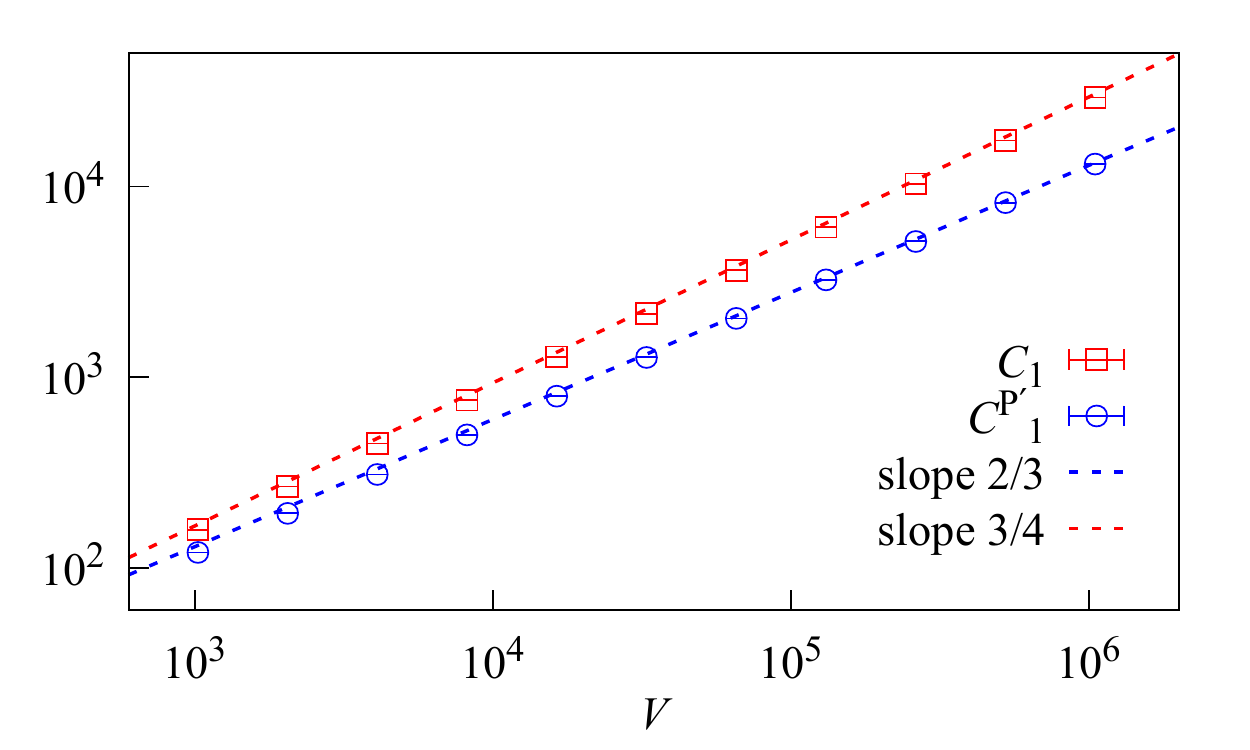}
    \caption{Log-log plot of the largest cluster size of the FK-Ising model $C_1$ and the largest cluster size of the configurations generated from the loop configurations where $\mathcal{F}_1 \leq 2V^{1/3}$, denoted as $C_1^{\rm P'}$, versus $V$. The red dashed line with slope $2/3$ and the blue dashed line with slope $3/4$ imply the difference between their volume fractal dimensions.}
    \label{fig:PerS_proof}
\end{figure}
To further verify our conjecture, we observe the scaling behavior of the largest FK cluster generated by performing the percolation process to the loop configurations where $\mathcal{F}_1\leq2V^{1/3}$, denoted as $C^{\rm P'}_1$. We show the data of $C^{\rm P'}_1$ and the largest cluster size of the FK Ising model $C_1$ in Fig.~\ref{fig:PerS_proof}; the former scales as $V^{2/3}$ and the latter scales $V^{3/4}$. This confirms our conjecture that the percolation sector in the FK Ising model corresponds to the loop configurations with the largest loop size of order $V^{1/3}$.

Thirdly, we consider the case away from  the critical point and define $t = (K_c-K)/K_c$.   When the critical point is approached  from high-temperature side ($t>0$), the magnetic susceptibility $\chi(t,V) = V \langle m^2 \rangle  = V^{2\yh-1} \tilde{\chi}(tV^{\yt})$ with renormalization-group exponents $(\yt=1/2, \yh=3/4)$. 
% Based on the FSS assumption, the scaling function $\tilde{\chi}(x) \sim \scrO(1)$ as $x\to 0$, where one would obtain the critical FSS behavior $\chi \sim \sqrt{V}$;
Based on the FSS assumption, as  $x \to \infty$, the scaling function $\tilde{\chi}(x) \sim x^{-\gamma}$ with $\gamma = (2\yh-1)/\yt=1$, which recovers the thermodynamic scaling behavior $\chi(t) \sim t^{-\gamma}$. 

%By the FSS assumption, the scaling function follows $\tilde{\chi}(x) \sim \scrO(1)$ as  $x \to 0$ and $\tilde{\chi}(x) \sim x^{-\gamma}$ as $x\to \infty$ with $\gamma = (2\yh-1)/\yt=1$. 
%Thus, one would expect $\tilde{\chi}(x) \sim x^{-\gamma}$ as $x \to \infty$ via the FSS assumption, where one recovers its thermaldynamic scaling behavior $\chi \sim |t|^{-\gamma}$ with the exponent $\gamma$ taking the mean-field prediciton $\gamma=1$.   
Recall in Sec.~\ref{theoretical analysis}, the bond number $B=  \frac{1}{2}V\langle m^2 \rangle + O(1) =\frac{1}{2} \chi + O(1)$. Thus, one would expect 
        \begin{equation}
            B = \sqrt{V} \tilde{B}(t\sqrt{V}), 
        \end{equation}
where the scaling function $\tilde{B} (x) \sim x^{-1}$ as $x \to \infty$. Then, if one takes $t=O(V^{-1/3})$, one would obtain $B \sim \sqrt{V} \cdot (V^{-1/3 + 1/2})^{-1} = V^{1/3}$, which is the same as the scaling of the bridge-free bond number for the CG-percolation model~\cite{PhysRevE.97.022107}. At this temperature, the FK clusters obtained by adding bonds via the LC algorithm are expected to behave the same as CG-percolation clusters, which explains the existence of percolation scaling window in the FK Ising model and the width is of order $O(V^{-1/3})$. While the temperature is decreased from $K_c$, the bond number increases and no percolation scaling window is observed.

% Now, we study the process of adding bond from the loop configuration with the probability near $p_c=\frac{1}{V-1}$. The result of Binder cumulant ratio $Q_s(p,V)$ are shown in Fig~\ref{fig:Qs1}. We can perform a fit with the data of $Q_s$:
% \begin{align}
%             r =&r_0+r_1(\frac{p}{p_c}-1)L^{y_r}+r_2(\frac{p}{p_c}-1)^2L^{2y_r} \notag\\ 
%   &+b_1L^{y_1}+b_2L^{y_2}+c_1(\frac{p}{p_c}-1)L^{y_1+y_r},
%   \label{fitQs}
% \end{align}
% where the renormalization relevant exponent $y_r$ is a characteristic exponent of the adding bond process. Figure.~\ref{fig:RQs} shows a collapse of the data with different system size indicating that the relevant exponent $y_r=\frac{1}{3}$ on the CG, which is consistent with the red-bond exponent. A precise simulation and fitting are needed in the further study.
\section{Discussion}
\label{secV}
In this work, we study the geometric properties of the complete-graph (CG) Ising model in the loop representation. Theoretically, we derive that the density of bonds decays as $V^{-1/2}$, which means the loop configurations are basically vacant in the thermodynamic limit. 
We numerically find that the volume fractal dimension for the first- and second-largest loop clusters is $1/2$, and the number of clusters scales as $\frac{1}{4}\ln{V}$. We also observe that the bond number is numerically consistent with the number of vertices in loop clusters, and this means these loops are uni-cyclic, which is similar to the bridge-free configurations of the CG-percolation model. 
Based on our numerical results, we conjecture the exact form of the probability distribution of the largest loop cluster and the cluster-number density $n(s,V)$. In Ref.~\cite{Ben-Naim2005Kinetica}, the authors used the rate equation approach to study and derive the cycle-length number density of the critical percolation on the CG, which scales as $(2sV)^{-1}$ with a cutoff at $O(V^{1/3})$. Thus, it has the same behavior as our $n(s,V)$ except the different cutoff.\par

The abundant critical behaviors in the Fortuin-Kasteleyn (FK) representation, i.e., the emergence of two length scales, two configuration sectors, and two scaling windows, are not found in the loop representation. But, via the loop-cluster (LC) joint model, results in the loop representation does provide a vivid and intuitive understanding to these critical behaviors in the FK representation. Under the LC joint model, the FK representation can be regarded as playing a percolation game on top of loop configurations. During the percolation process, almost all loops are connected together and end up with forming the largest FK cluster. Other FK clusters are basically these newly generated percolation clusters. \par

%Our results about the geometric properties of the loop representation reveal that the LC process is actually a percolation process. The emergence of two length scales in the FK representation is naturally shown as a result of the physical process where all loops with the same scaling behavior are merged in the largest FK cluster. 
%Other clusters in the FK representation are generated in the percolation-like process. In the thermodynamic limit, we can neglect the largest FK cluster, since its size is not extensive. Therefore, we conclude that the configuration space of the FK-Ising model is nearly equivalent to the one of percolation.  
%Meanwhile, we find that the vanishing percolation sector in the FK representation corresponds to the counterpart with $\mathcal{F}_1\leq \scrO(V^{1/3})$ of the loop representation.
%, which is related to the percolation scaling window ~\cite{luczak2006phase} and the size of bridge-free percolation clusters~\cite{PhysRevE.97.022107}.

It is generally believed that the CG is a mean-field approximation to high-dimensional tori. Recently, the FK Ising model on lattices above the upper critical dimension $d_c = 4$ was studied and the similar scaling behaviors (two length scales, two sectors and two scaling windows) were again observed~\cite{Fang_2022,Fang_2022_2}. More interestingly, in addition to the well-known upper critical dimension $d_c = 4$, these anomalous scaling behaviors uncover a new upper critical dimension $d_p = 6$, which cannot be observed in the spin representation. Therefore, a number of questions naturally arise. First, can the LC joint model provide an understanding to the anomalous behaviors of the FK representation on high-dimensional lattices? Second, can the two-upper-critical-dimensional phenomena be observed in the loop representation? Third, can the loop representation provide a straightforward understanding for the existence of two upper critical dimensions? These open questions will be investigated in our future work.

\section*{Acknowledgements}
This work has been supported by the National Natural Science Foundation of China (under Grant No. 12275263), the National Key R\&D Program of China
(under Grant No. 2018YFA0306501). We thank Pengcheng Hou and Tianning Xiao for valuable discussions.

\bibliographystyle{apsrev4-2}
%\bibliography{main}
%

\end{document}